%% file: 6dgenrs.tex
\documentstyle[12pt,psfig]{article}

\begin{document}

\begin{flushright}
hep-th/0401207
\end{flushright}

\vspace{5mm}

\begin{center}
{\Large \bf Scalar Hair of Global Defect and Black Brane World}\\[12mm]
Yoonbai Kim\\
{\it BK21 Physics Research Division and Institute of Basic Science,\\
Sungkyunkwan University, Suwon 440-746, Korea}\\
{\tt yoonbai$@$skku.edu}\\[5mm]
Dong Hyun Park\\
{\it Center for Proton Therapy, National Cancer Center\\
Goyang 411-769, Korea}\\
{\tt donghyun$@$ncc.re.kr}
\end{center}

\vspace{5mm}

\begin{abstract}
We consider a complex scalar field in $(p+3)$-dimensional bulk with a 
negative cosmological constant and study global vortices in two 
extra-dimensions. We reexamine carefully the coupled scalar and Einstein 
equations, and show that the boundary value of scalar amplitude at infinity 
of the extra-dimensions should be smaller than vacuum expectation value.
The brane world has a cigar-like geometry with an exponentially decaying 
warp factor and a flat thick $p$-brane is embedded. 
Since a coordinate transformation identifies the obtained brane world 
as a black $p$-brane world bounded by a horizon, 
this strange boundary condition of the scalar amplitude is understood 
as existence of a short scalar hair. 
\end{abstract}

{\it{Keywords}}: Global vortex, brane world

\newpage

\setcounter{equation}{0}
\section{Introduction}

In the recent developments, two main streams of brane world scenario have 
been proposed: one with large extra-dimensions by 
Arkani-Hamed-Dimopoulos-Dvali (ADD)~\cite{Arkani-Hamed:1998rs},
and the other with a warp geometry by 
Randall-Sundrum (RS)~\cite{Randall:1999ee,Randall:1999vf}.

In Ref.~\cite{Randall:1999vf}, RS considered a flat 3-brane in 
five-dimensional anti-de Sitter (AdS) bulk spacetime with one 
extra-dimension.
Newtonian gravity was reproduced on the 3-brane, in its sufficiently
low energy limit. Subsequently numerous related topics have been explored,
and, among them,
two natural extensions have been the brane world with two
extra-dimensions~\cite{Chodos:1999zt}
and construction of thick brane world~\cite{DeWolfe:1999cp}--\cite{Kim:2004vt}.

In this paper, we revisit the brane world of warp geometry 
composed of a flat thick $p$-brane
and two extra-dimensions with rotational symmetry.
Before considering bulk fields to make the brane thick, we examine
Einstein equations by assuming only a negative
bulk cosmological constant, and obtain general static vacuum solutions under 
the ansatz of a metric with warp factor. 
Those solutions are characterized mainly by
an arbitrary integration constant, and the geometry with an exponentially
decaying warp factor is uniquely given only when a particular value of that 
parameter is chosen. The obtained unique brane world from a vacuum solution is 
identified with
that of AdS branch in Ref.~\cite{Gregory:1999gv,Gregory:2002tp}.
It is also confirmed that any black $p$-brane structure cannot be formed
with a horizon at a finite radius.

Suppose that we have 
a complex scalar field with global U(1) symmetry in the bulk.
Then global vortices, thick global $p$-branes in our context, may be formed 
due to the topological winding between the broken vacua and
spatial infinity of the two extra-dimensions. Since these global defects
are gravitating without cosmological constant, a physical singularity is 
unavoidable~\cite{Cohen:1988sg}
and does not disappear for the cases of nonzero $p$~\cite{Cohen:1999ia}.
Once a negative cosmological constant is added, then a logarithmically
divergent energy becomes not so harmful because of the negative vacuum 
energy proportional to its spatial volume. The corresponding
spacetime is free from the physical singularity but can support 
a charged Ba\~{n}ados-Teitelboim-Zanelli (BTZ) black hole 
structure~\cite{Kim:1997ba}. Since the
brane world with a warp geometry along this line means an extension 
from 0-brane to general flat $p$-brane, appearance of both a 
singularity-free brane world with an exponentially decaying warp 
factor~\cite{Gregory:1999gv,Olasagasti:2000gx,Gherghetta:2000jf,Gregory:2002tp}
and structure of a black brane 
world~\cite{Moon:2000hn,Moon:2001ck} is naturally understood.
A noteworthy observation is the fact that the obtained brane world in terms of
the warp metric coincides exactly with the interior of the black brane world
bounded by the degenerated horizon. Since our brane world is a static 
patch, an intriguing question arises about
the existence of a short scalar hair. Specifically, the question is 
whether or not the amplitude of a bulk complex scalar field can have the vacuum
expectation value as its boundary condition at spatial infinity of the 
extra-dimensions.
By analyzing the coupled field equations, we will show that the answer is 
negative and then the scalar amplitude at spatial infinity has a smaller 
value than the 
vacuum expectation value. In the context of black $p$-brane world, this strange 
boundary condition of the scalar amplitude is understood as existence of a 
short scalar hair at the degenerated horizon. In order to have
vanishing scalar hair, either the limit of vanishing scalar potential 
or that of vanishing cosmological constant should be taken, but both of them 
are unwanted limits. We will also explain the existence 
of an extremal
black $p$-brane through a Hodge duality between derivative term of
the scalar phase and a dual five-form field strength tensor.

In section 2 we consider a bulk composed of a flat $p$-brane and rotationally
symmetric two extra-dimensions, and obtain all the static AdS
solutions under the metric with a warp factor. In section 3 a bulk complex
scalar field is assumed and show that the boundary value of the scalar field 
for a global defect cannot arrive at the vacuum expectation value 
at infinity of the extra-dimensions.
In section 4 we identify the boundary condition as the existence of
a short scalar hair around the degenerated horizon of an extremal black
brane world through a coordinate transformation. We conclude in section 5 
with a few comments and discussions.

\setcounter{equation}{0}
\section{Vacuum Solutions and Warp Geometry}

In this section, we revisit the static warp geometry of a flat $p$-brane 
configuration in $D$-dimensional bulk spacetime. 
The extra-dimensions are two and no
bulk matter is assumed except for the gravity with a negative bulk 
cosmological constant.
We will obtain all the static vacuum solutions under the metric with warp
factor, which are characterized by one integration constant, and will show 
that an exponentially decaying warp factor is obtained only when
this parameter is chosen by a specific value.

We begin with the Einstein-Hilbert action in $(D=p+3)$-dimensions
with a nonpositive bulk cosmological constant $\Lambda$
\begin{equation}\label{ehac}
S_{{\rm EH}}=\int d^{D}x\sqrt{-g_{D}}\left[-\frac{M^{p+1}_*}{16\pi}
(R+2\Lambda)\right],
\end{equation}
where $M_*$ is the fundamental scale of higher dimensional gravity.
The following ansatz for the metric is adopted for description of
the warp geometry with convenience
\begin{eqnarray}
ds^2 &=& g_{AB}dx^{A}dx^{B}\nonumber\\
&=& e^{2A(r)} dx^{\mu}dx_{\mu} - dr^2 - C^2(r) d\theta^2 ,
\label{metric}
\end{eqnarray}
where indices, $A, B, ...$, denote those of $D$-dimensional bulk
and $\mu$ stands for spacetime coordinates of the flat $p$-brane. 

Einstein equations from the action (\ref{ehac}) are simplified as
\begin{equation}\label{imeq1}
-\frac{C^{''}}{C}-pA^{''}+A^{'}\frac{C^{'}}{C}=0,
\end{equation}
\begin{equation}\label{imeq2}
A^{''}+A^{'}\frac{C^{'}}{C}+(p+1)A^{' 2}
=\frac{2|\Lambda|}{p+1},
\end{equation}
\begin{equation}\label{imeq3}
A^{''}-\frac{C^{''}}{C}-pA^{'}\frac{C^{'}}{C}+(p+1)A^{'2}=0,
\end{equation}
and they reduce to two first-order equations
\begin{eqnarray}
\left(A^{'}-\frac{C^{'}}{C}\right)Ce^{(p+1)A}=\kappa , \label{vaceq1}\\
\frac{p}{2}A^{' 2}+A^{'}\frac{C^{'}}{C}=-\frac{\Lambda}{p+1}, \label{vaceq2}
\end{eqnarray}
where $\kappa$ is an integration constant. 
In the remaining part of this section, let us examine the Einstein equations
(\ref{vaceq1})--(\ref{vaceq2}) and show that the unique solution
of $\kappa=0$ among a set of infinite AdS solutions can depict
RS geometry with an exponentially decreasing warp factor.

When $\kappa$ is zero, the first equation (\ref{vaceq1}) gives $C=\alpha
e^{A}$. Substituting it into the second equation (\ref{vaceq2}), we have a set
of exact solutions which are all the static solutions of vanishing $\kappa$
\begin{eqnarray}
A_{\pm}(r)&=&\pm \frac{2\omega}{p+2}~r+\beta_{\pm},
\label{pasol}\\
C_{\pm}(r)&=&\alpha_{\pm}e^{A_{\pm}(r)}=\alpha_{\pm}
e^{\pm\frac{2\omega}{p+2}r+
\beta_{\pm}},
\label{pcsol}
\end{eqnarray}
where $\alpha_{\pm}$ and $\beta_{\pm}$ are integration constants and 
\begin{equation}\label{omega}
\omega=\sqrt{\frac{(p+2)|\Lambda|}{2(p+1)}}.
\end{equation}
Note that $\beta_{\pm}$ can be fixed to be zero with the aid of 
reparametrization of the coordinates $x^{\mu}$. The solution
$(A_{-},C_{-})$ describes an exponentially decaying warp factor in front 
of both spacetime metric of the $p$-brane and angle variable $\theta$ of 
the extra-dimensions. Therefore, a neck of the two extra-dimensions 
becomes thin as radius $r$ increases. It is a $p$-dimensional candidate of 
brane world of the RS type~\cite{Gregory:1999gv}.

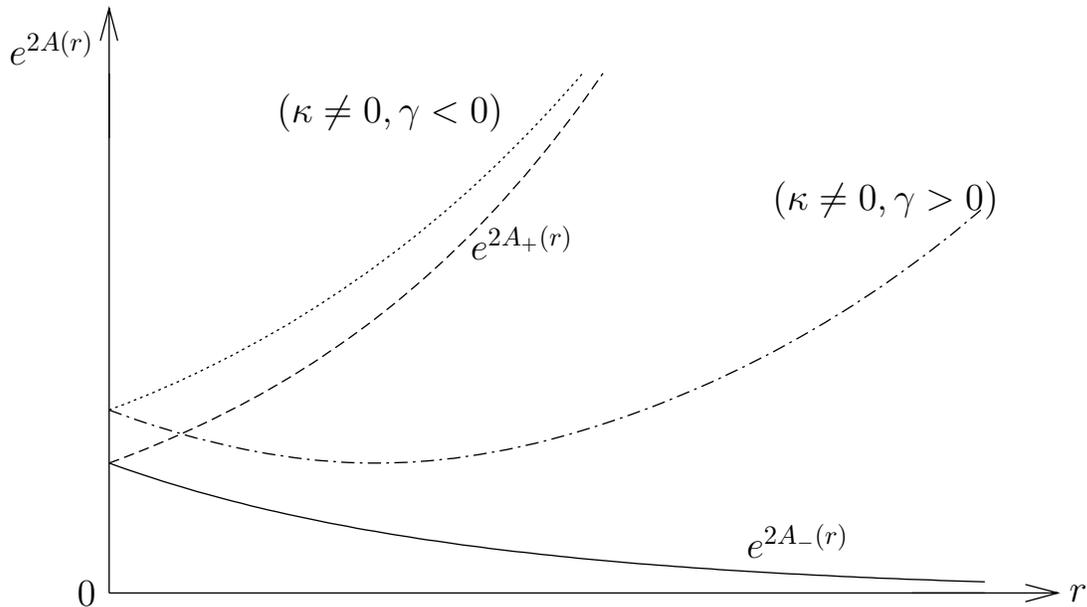
\begin{figure}[ht]
\begin{center}
\input{6dfig1}
\end{center}
\caption{Four representative shapes of the metric function $e^{2A(r)}$:
solid line for $e^{2A_{-}}$, dashed line for $e^{2A_{+}}$, and both dotted
$(\gamma<0)$ and dashed-dotted $(\gamma>0)$ lines for the $\kappa\ne 0$
solutions.}
\label{6dfig1}
\end{figure}

When $\kappa\ne 0$, general solutions of the Einstein equations
(\ref{simeq1})--(\ref{simeq3}) are also obtained as 
\begin{eqnarray}
A(r)&=&\frac{2}{p+2}\ln\left[\cosh\left(
\omega~r-\gamma\right)\right]+\delta, 
\label{areq}\\
C(r)&=& -e^{\delta}\frac{\kappa}{\omega}\left[\cosh\left(
\omega~r-\gamma\right)\right]^{-\frac{p}{p+2}}
\sinh\left(\omega~r -\gamma\right) ,
\label{creq}
\end{eqnarray}
where $\gamma$ and $\delta$ are integration constants.
Again, $\delta$ can be fixed to be zero with the help of reparametrization
of the coordinates $x^{\mu}$. A noteworthy property of the
solutions is that the $\kappa\ne 0$ solutions are disconnected from the
$\kappa=0$ solutions, i.e., $\kappa\rightarrow 0$ limit of the $\kappa \ne
0$ solutions (\ref{areq})--(\ref{creq}) do not coincide with the $\kappa=0$
solutions (\ref{pasol})--(\ref{pcsol}). If we rewrite the solutions
(\ref{areq})--(\ref{creq}) by using exponential functions;
\begin{eqnarray}
A(r)&=&\frac{2}{p+2}\ln (\gamma_{+}e^{\omega r} +\gamma_{-}e^{-\omega r}),
\\
C(r)&=&-\frac{\kappa}{\omega}(\gamma_{+}e^{\omega r} +\gamma_{-}e^{-\omega r}
)^{-\frac{2}{p+2}}(\gamma_{+}e^{\omega r} -\gamma_{-}e^{-\omega r}),
\end{eqnarray}
where $\gamma_{\pm}\equiv e^{\mp \gamma+\frac{p+2}{2}\delta-\ln 2}$. One
may think that a solution with an exponentially decaying warp factor is
obtained by taking the limit $\gamma_{+}=0$. However, when $\gamma_{+}$
is taken to be zero, the definition of $\gamma_{+}$ asks either $\gamma$
or $-\delta$ to diverge to positive infinity so that $\gamma_{-}$ becomes
divergent or vanishes, respectively.
Finally we only have either an unwanted solution 
$(\gamma_{+}=0,\gamma_{-}=\infty )$ or a trivial solution
$(\gamma_{+}=0, \gamma_{-}=0)$. The disconnectedness of the solutions
is easily understood by appearance of 
a formal symmetry property for vanishing $\kappa$,
that the first-order equations (\ref{vaceq1})--(\ref{vaceq2}) possess 
a formal parity, $r\rightarrow -r$ as far as $A^{'}$ and $(\ln C)^{'}$ 
share the same parity. As shown in Fig.~\ref{6dfig1}, the $\kappa\ne 0$ 
solution
has minimum value $e^{2A(r_{\rm min})} =e^{2\delta}(\cosh \gamma)^{2/(p+2)}$
at $r_{\rm min}=0$ when $\gamma$ is negative (see the dotted line in 
Fig.~\ref{6dfig1}), but it has a minimum $e^{2A(r_{\rm min})}=e^{2 \delta}$ 
at $r_{\min}=\gamma/\omega$ (see the dashed-dotted line in 
Fig.~\ref{6dfig1}),
when $\gamma$ is positive. For sufficiently large $r$, both metric functions 
$e^{A(r)}$ and $C(r)$ asymptote exponentially growing solutions 
irrespective of value of any nonvanishing $\kappa$. 

In conclusion, one of the $\kappa=0$ solutions, 
$A_{-}(r)$ and $C_{-}(r)$ in Eqs.~(\ref{pasol})--(\ref{pcsol}) in 
Ref.~\cite{Gregory:1999gv} (see the solid line in Fig.~\ref{6dfig1}), 
is the unique static vacuum solution to describe RS type brane
world with an exponentially decaying warp factor. Since geometry of our
interest is given by a fine tuned solution but all the other static
vacuum solutions have totally different asymptotic behavior, almost all
the new solutions are likely to be connected with exponentially growing
solutions. 

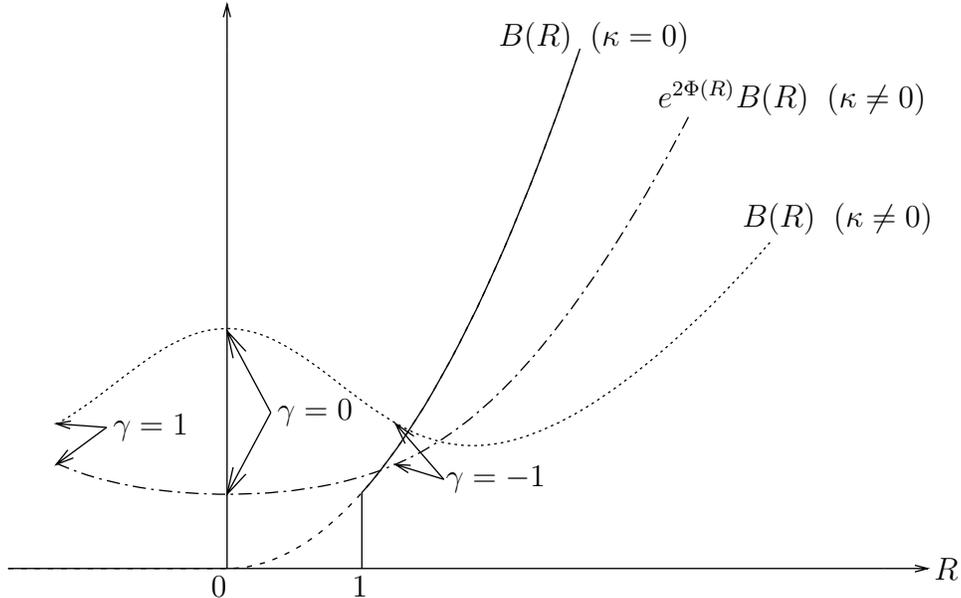
\begin{figure}[h]
\begin{center}
\input{6dfig2}
\end{center}
\caption{Plot of the various metric functions as functions of the new
radial coordinate $R$: $B$ (solid and dashed line) for the $\kappa=0$
solutions, and $e^{2\Phi}B$ (dotted line) and $B$ (dashed-dotted line) for
the $\kappa\ne 0$
solutions. The points indicated by arrows on the curves stand for minimum
values of $R$ for given values of $\gamma$.}
\label{6dfig2}
\end{figure}

A familiar coordinate for AdS spacetime is Poincar\'{e} 
coordinate
\begin{equation}\label{metric1}
ds^{ 2}=e^{2\Phi (R)}B(R) g_{\mu\nu}dx^\mu dx^\nu-\frac{{dR}^2}{B(R)}
-R^{2}d\theta^2 .
\end{equation}
Throughout a coordinate transformation $R=2\omega C(r)/(p+2)$ with the
help of reparametrization of the $p$-brane coordinates $x^{\mu}/\beta_{\pm}
\rightarrow x^{\mu}$ $(\mu=0,1,\cdots,p)$,
the vacuum solutions of the warp factor $(A_{\pm},C_{\pm})$ 
in Eqs.~(\ref{pasol})--(\ref{pcsol}) are
transformed to a well-known form of the metric such as $\Phi(R)=0$ and
$B(R)=R^{2}$. Since the $(+)$-solution corresponds to a patch of range
$1\le R\le\infty$ (see solid line in Fig.~\ref{6dfig2}) and the 
$(-)$-solution to that of range $0\le R\le 1$ (see dashed line in
Fig.~\ref{6dfig2})
under the coordinate transformation, spatial infinity $(r=\infty)$ of the
$(-)$-solution is
translated into a horizon at the origin $(R=0)$. For a $0$-brane case, it 
is nothing but the horizon of a BTZ black hole at its zero mass limit.
 
Let us try a coordinate transformation from the warp coordinate
(\ref{metric}) to the Poincar\'{e} coordinate (\ref{metric1}) for the 
general solutions (\ref{areq})--(\ref{creq}). Comparison between the two
metrics (\ref{metric}) and (\ref{metric1}) provides
\begin{eqnarray}
R&=&\frac{2\omega}{p+2}C(r)\longrightarrow \left(e^{-\delta}
\frac{p+2}{2\kappa}R\right)^{2}=x^{-\frac{2p}{p+2}}(x^{2}-1),
\label{com1}\\
B(R)&=&\left(\frac{dC}{dr}\right)^{2}=
\frac{x^{2}}{x^{2}-1}\left(1+\frac{p}{2x^{2}}\right)^{2}R^{2},
\label{com2}\\
\Phi(R)&=&A(r)-\ln \frac{p+2}{2\omega}\sqrt{B(R)}=
\ln\left(e^{\delta}\frac{2\omega}{p+2}\frac{x^{\frac{p+4}{p+2}}
\sqrt{x^{2}-1}}{x^{2}+p/2}\frac{1}{R}\right),
\label{com3}
\end{eqnarray}
where $x=\cosh (\omega r-\gamma)$. If we rewrite Eq.~(\ref{com1}) in terms of 
a new variable $y\equiv x^{\frac{2}{p+2}}$, then we arrive at an algebraic
equation
\begin{equation}
y^{p+2}-\left(e^{-\delta}\frac{p+2}{2\kappa}R\right)^{2}y^{p}-1=0.
\end{equation}
When $p=3$, it is obvious that any analytic form of a particular solution
has not been reported from the above fifth-order algebraic equation
except for the trivial case of $R=0$. Therefore, the explicit
coordinate transformation to the Poincar\'{e} coordinate cannot be 
obtained for the general $\kappa \ne 0$ vacuum solution
(\ref{areq})--(\ref{creq}). 
It may imply impossibility to solve the general solutions including
Eqs.~(\ref{areq}) and (\ref{creq}) as a patch in the Poincar\'{e}
coordinates (\ref{metric1}) though we obtained those exact solutions in the
warp coordinates (\ref{metric}).
A rescaling of the spacetime variables of the $p$-brane as $2\omega
e^{\delta}x^{\mu}/(p+2)\rightarrow \tilde{x}^{\mu}$ leads to
\begin{equation}
ds^{2}=x^{\frac{4}{p+2}}\eta_{\mu\nu}d\tilde{x}^{\mu}d\tilde{x}^{\nu}
-\frac{x^{2}(x^{2}-1)}{(x^{2}+p/2)^{2}}\frac{dR^{2}}{R^{2}}-R^{2}d\theta^{2}.
\end{equation}
Radial coordinate $R(r)$ is a monotonically-increasing function of $r$ 
with the range
\begin{equation}
e^{\delta}\frac{2\kappa}{p+1}(\cosh \gamma)^{-\frac{p}{p+2}}\sinh \gamma
\le R \le \infty ,
\end{equation}
where the minimum value of $R$ is negative, zero, or positive according to 
that of $\gamma$. Since $1\le \cosh \gamma\le x\le \infty$, $\Phi(R)$ is finite 
and $B(R)$ is positive for every $R$, which means that no horizon structure 
can be realized from all the $\kappa\ne 0$ vacuum solutions as shown
in Fig.~\ref{6dfig2}.

\setcounter{equation}{0}
\section{Global Defect with Fine Tuning}
We showed that the brane world geometry with an exponentially decaying
warp factor could only be obtained in $(p+3)$-dimensions 
as a static vacuum solution of a specific integration 
constant (\ref{pasol})$\sim$(\ref{pcsol}), and all the other solutions are 
those with an exponentially growing warp factor (\ref{areq})$\sim$(\ref{creq}). 
In this section we will examine smooth global U(1) vortices and 
address a question whether or not such defect configurations form a new 
brane world with an exponentially decaying warp factor at its asymptote.

A gravitating global defect in two extra-dimensions 
is a global vortex solution of a field theory coupled to
gravity with a spontaneously broken continuous global symmetry. A
prototypical model has U(1) symmetry and a `Mexican hat' potential:
\begin{equation}\label{ac}
S=\int d^{D}x\sqrt{-g_{D}}\left[-\frac{M^{p+1}_*}{16\pi}(R+2\Lambda)+
\frac12 \nabla_A\bar{\phi} \nabla^A \phi - 
\frac{\lambda}{4} ( \bar{\phi}\phi -v^2)^2 \right].
\end{equation}
Here we consider a static global vortex in two spatial extra-dimensions, 
and its transverse space has $p$-dimensions - 
$p$-brane configurations in the spacetime of dimensions $D=p+3$. 
We shall call this $p$-dimensional defect as {\it global $p$-brane}
in what follows.

When $n$ vortices in two
extra-dimensions are superimposed at the origin, an ansatz for the global
$p$-brane is
\begin{equation}\label{ansa}
\phi = v f(r) e^{in\theta}.
\end{equation}
Note that both the complex scalar field $\phi$ and its vacuum expectation 
value $v$ has a canonical mass dimension of $(p+1)/2$. To be a nonsingular
configuration, the scalar amplitude $f$ should vanish at the origin, i.e.,
$f(0)=0$ is a boundary condition.

In terms of dimensionless variables and quantities in the metric (\ref{metric})
and the action (\ref{ac})
\begin{equation}\label{resc}
\sqrt{\lambda} v x^{\mu}\rightarrow x^{\mu}~(\sqrt{\lambda} v r
\rightarrow r),~~
\sqrt{\lambda} v C\rightarrow C,~~\Lambda/\lambda v^{2}\rightarrow \Lambda,
~~G_{D}=v^{2}/M_{\ast}^{p+1},
\end{equation}
scalar field equation is
\begin{equation}\label{scaleq}
f^{''}+\left[\frac{C^{'}}{C}+(p+1)A^{'}\right]f^{'}
-\left[\frac{n^2}{C^2}+(f^2-1)\right]f=0.
\end{equation}
Einstein equations are simplified as
\begin{equation}\label{simeq1}
-\frac{C^{''}}{C}-pA^{''}+A^{'}\frac{C^{'}}{C}=
8 \pi G_D f^{' 2},
\end{equation}
\begin{equation}\label{simeq2}
A^{''}+A^{'}\frac{C^{'}}{C}+(p+1)A^{' 2}
=\frac{2}{p+1}\left[|\Lambda|-2\pi G_{D}(f^{2}-1)^{2}\right],
\end{equation}
\begin{equation}\label{simeq3}
A^{''}-\frac{C^{''}}{C}-pA^{'}\frac{C^{'}}{C}+(p+1)A^{'2}
=8\pi G_{D} \frac{n^{2}}{C^{2}}f^{2}.
\end{equation}
One may notice that no explicit metric dependence on the right-hand side 
of both equations (\ref{simeq1})--(\ref{simeq2}).
In Eq.~(\ref{simeq2}), the effective cosmological constant $\lambda v^{2}
\left[-|\Lambda| +2 \pi G_{D} (f^2-1)^2\right]$ appears instead of 
the cosmological constant $\lambda v^{2}\Lambda$.

First of all, let us investigate the core region of global vortex, 
where the scalar amplitude $f(r)$ grows as the radius increases.
Up to now  we find no exact solution from the coupled equations of  
motion (\ref{scaleq})--(\ref{simeq3}), we employ two  
approaches to obtain an approximate solution near the origin: One is to use
power series solution of the exact equations near the origin and the
other is to solve approximated equations exactly. 
Then we will compare both results.
Since the scalar amplitude of a global vortex lives
near the symmetric vacuum at its core, i.e., $f \approx 0$, forms of the 
approximated equations from Eqs.~(\ref{scaleq})--(\ref{simeq3}) become the
same as those of the vacuum (\ref{vaceq1})--(\ref{vaceq2}) 
except for the change 
of an effective cosmological constant $2\pi G_{D}+\Lambda$ instead 
of $\Lambda$, which can have any signature. Therefore, we can use the exact 
solutions, Eqs.~(\ref{pasol})--(\ref{creq}):
One with vanishing $\kappa$ is
\begin{eqnarray}
A_{\pm}(r)&=&\pm \frac{2\tilde{\omega}}{p+2}~r+\tilde{\beta}_{\pm},
\label{pasolt}\\
C_{\pm}(r)&=&\tilde{\alpha}_{\pm}e^{A_{\pm}(r)}=\tilde{\alpha}_{\pm}
e^{\pm\frac{2\tilde{\omega}}{p+2}r+ \tilde{\beta}_{\pm}},
\label{pcsolt}
\end{eqnarray}
where $\tilde{\alpha}_{\pm}$ and $\tilde{\beta}_{\pm}$ are integration 
constants and 
\begin{equation}\label{omt}
\tilde{\omega} =\sqrt{\frac{(p+2)(2\pi G_D+\Lambda)}
{2(p+1)}}.
\end{equation}
Here $\tilde{\beta}_{\pm}$ can also set to be zero by a reparametrization 
of the $p$-brane spacetime coordinates $x^\mu$. The other is general 
solutions of the nonvanishing $\kappa$
\begin{eqnarray}
A(r)&=&\frac{2}{p+2}\ln\left[\cos\left(\tilde{\omega} r-\tilde{\gamma}\right)
\right]+\tilde{\delta} 
\label{acore}\\
&\approx&\left(\frac{2}{p+2}\ln\cos \tilde{\gamma}+\tilde{\delta}\right)+
\frac{2\tilde{\omega}}{p+2}(\tan \tilde{\gamma}) r
-\frac{\tilde{\omega}^{2}}{p+2}(1-2\tan^{2}\tilde{\gamma})r^{2}+\cdots ,
\label{acore0}\\
C(r)&=&- e^{-(p+1){\tilde{\delta}}} \frac{\kappa}{\tilde{\omega}}
\left[\cos\left(\tilde{\omega} r-\tilde{\gamma}\right) 
\right]^{-\frac{p}{p+2}} \sin\left(\tilde{\omega}r -\tilde{\gamma}\right) 
\label{ccore}\\
&\approx&e^{-(p+1)\tilde{\delta}}\frac{\kappa}{\tilde{\omega}}\sin 
\tilde{\gamma}(\cos \tilde{\gamma})^{-\frac{p}{p+2}} 
\left[1-\left(\frac{p}{p+2}\tan \tilde{\gamma}-\cot \tilde{\gamma}\right)
\tilde{\omega}r \right.
\nonumber\\
&&\left. +(p-1)\left(\frac{1}{p+2}+\frac{p}{2}\tan^{2}\tilde{\gamma}\right)
\tilde{\omega}^{2} r^{2}\right],
\label{ccore0}
\end{eqnarray}
where $\tilde{\gamma}$ and $\tilde{\delta}$ are integration constants of which 
$\tilde{\delta}$ can again be fixed to be zero by a reparametrization 
of the $p$-brane spacetime coordinates $x^\mu$. When the effective 
cosmological constant $2 \pi G_D + \Lambda$ is positive, exact 
solutions of the approximated equations (\ref{pasolt})--(\ref{ccore0}) are
expressed by trigonometric functions instead of hyperbolic functions.

On the other hand, if we take into account the scalar field $f$, 
then series solutions near the origin are given from 
Eqs.~(\ref{scaleq})--(\ref{simeq3})
\begin{eqnarray}
f(r)&\approx& f_{0}r^{n/c_{0}},\label{f0r}\\
A(r)&\approx& a_{0} - \frac{{\tilde{\omega}}^2}{p+2}r^2 ,
\label{aexp0}\\
C(r)&\approx& c_{0}r\left\{ 1 + \frac{1}{6} \left[ \frac{2(p-1)
{\tilde{\omega}}^2}{p+2} - f_{0}^{2} \delta_{\frac{n}{c_{0}}1} \right]
r^{2} \right\} ,
\label{c0r}
\end{eqnarray}
where $f_{0}$, $a_{0}$, $c_{0}$ are undetermined constants determined by
proper behavior at the opposite boundary. $c_{0}$ should be restricted to make
$n/c_{0}$ a natural number for regularity and $a_{0}$ can be absorbed by a
reparametrization of the $p$-brane spacetime coordinates $x^{\mu}$. 
Comparison of
Eq.~(\ref{aexp0}) with Eq.~(\ref{pasolt}) leads obviously to the conclusion
that the $\kappa=0$
solution is not consistent with any possible nonsingular vortex
configuration. On the other hand, expansion of the $\kappa\ne 0$ solution
(\ref{ccore0}) can be consistent with Eq.~(\ref{aexp0}) if $\tilde{\gamma}$ is
zero, $\tilde{\delta}$ is identified with $a_{0}$, and
$\kappa=-\frac{\tilde{\omega}}{(p+1)a_{0}}\ln(c_{0}/\tilde{\omega})$. 
Even if we take into account possible effect of the scalar amplitude $f$ and
vorticity $n$ outside the vortex core, it seems unlikely to find 
a configuration
of global vortex of which asymptotic geometry is smoothly connected to the
brane world of an exponentially decaying warp factor, $A_{-}(r)$ and 
$C_{-}(r)$ in Eqs.~(\ref{pasol})--(\ref{pcsol}), given as an AdS
vacuum solution. According to the above argument, even if we obtain 
a geometry of an exponentially decaying warp factor at asymptotes 
with the global vortices, it is irrelevant with that from the vacuum 
solution (\ref{pasol})--(\ref{pcsol}). We 
will show that it is indeed the case.

Once any brane world is assumed to be formed from a bulk global U(1) defect 
solution of a complex scalar theory, natural boundary condition for the
scalar amplitude is to reach vacuum expectation value at spatial infinity,
$f(\infty)=1$, without doubt~\cite{Gregory:1999gv,Olasagasti:2000gx,
Gherghetta:2000jf,Moon:2000hn,Moon:2001ck,Gregory:2002tp}. 
From here on, let us read possible sets of boundary conditions 
for the metric functions
$A(r)$ and $C(r)$ at spatial infinity when $\mathop{\lim}
\limits_{r\to \infty}f(r)=1$, and show that $f(\infty)=1$ cannot be taken 
in order to obtain a new brane world geometry with an exponentially
decaying warp factor formed by a global $p$-brane. 

We consider two cases
categorized by asking a criterion that whether or not the topological
term $n^{2}f/C^{2}$ in the scalar equation (\ref{scaleq}) is involved
in determining the boundary conditions of the metric functions $A(r)$ and
$C(r)$. 
First, if it is not the case, then the negative cosmological term
dominates and the boundary conditions for
$A(r)$ and $C(r)$ at spatial infinity reduce obviously to those of the pure
broken vacuum so that asymptotic solutions should be consistent with either 
geometry of the exponentially decaying warp factor, $A_{-}(r)$ and $C_{-}(r)$
in Eqs.~(\ref{pasol})--(\ref{pcsol}), or a hyperboloid, 
$A_{+}(r)$ and $C_{+}(r)$
in Eqs.~(\ref{pasol})--(\ref{pcsol}) and $A(r)$ and $C(r)$ in 
Eqs.~(\ref{areq})--(\ref{creq}). Second, when the topological term proportional
to $n^{2}$ contributes to determination of the boundary conditions
of the metric functions, which is specifically $T^{t}_{\;t}=
T^{z_{i}}_{\; z_{i}}=T^{r}_{\;r}=-T^{\theta}_{\;\theta}=n^{2}/2C^{2}$
at spatial infinity. 
For the boundary configuration of the scalar field, $f(\infty)=0$ and $n\ne 0$,
we obtain resultant Einstein equations from 
Eqs.~(\ref{simeq2})--(\ref{simeq3}):
\begin{equation}\label{nein1}
A^{''}+A^{'}\frac{C^{'}}{C}+(p+1)A^{' 2}=-\frac{2|\Lambda|}{p+1},
\end{equation}
\begin{equation}\label{nein3}
A^{''}-\frac{C^{''}}{C}-pA^{'}\frac{C^{'}}{C}+(p+1)A^{'2}
=8\pi G_{D} \frac{n^{2}}{C^{2}}.
\end{equation}
If we are interested in neither a coordinate singularity nor a geometry
of a power law warp factor such as $A^{\prime}(r) \sim r^m$ where $m$ is a
natural number, then consistency with the equations 
(\ref{nein1})--(\ref{nein3})
allows only minimal warp geometry such as $\mathop{\lim}\limits_{r\to \infty} 
A^{'} =$ constant irrespective of functional form of the metric $C(r)$. Then 
one more set of boundary conditions is easily obtained by a simple solution 
of the above equations at infinity~\cite{Gregory:1999gv}:
\begin{eqnarray}
A^{'}(\infty)&=&\pm \frac{2 \omega}{\sqrt{p+2}},
\label{aGre}\\
C(\infty)&=&\pm\sqrt{2 \pi G_D (p+2)}\,\frac{n}{\omega}.
\label{cGre}
\end{eqnarray}
The obtained boundary conditions (\ref{aGre})--(\ref{cGre}) are consistent with
the remaining Einstein equation (\ref{simeq1}). A noteworthy character of this
geometry is nonvanishing radius of the extra-dimensions represented by 
nonzero boundary value of the metric function $C$ (\ref{cGre}). Therefore, 
asymptotic region of the bulk spacetime has a finite size throat of a
cylinder. 

As the last test let us analyze asymptotic behavior of the
scalar field, represented by small perturbation $\delta f(r)$ defined by 
$f(r) \approx 1 
- \delta f(r)$ for sufficiently-large $r$.
Then the scalar equation (\ref{scaleq}) is approximated as 
\begin{equation}\label{delf}
\delta f^{''} + \left[ \frac{C^{'}}{C} + (p+1) A^{'} \right] \delta f^{'}
- \left( \frac{n^2}{C^2} + 2 \right) \delta f \approx -\frac{n^2}{C^2},
\end{equation}
where smallness of the derivatives, $|\delta f^{'}|\sim|\delta f^{''}|\ll 1$, 
has also been forced because of smoothness of the scalar field. For the brane
world of an exponentially decaying warp factor $A_{-}(r)$ and throat
$C_{-}(r)$ in Eqs.~(\ref{pasol})--(\ref{pcsol}), the leading behavior of
the right-hand side of Eq.~(\ref{delf}) forbids smallness of $\delta f$ in its 
left-hand side, i.e., $-n^2 \delta f(r)/C^2\approx -n^2/C^2$ in 
Eq.~(\ref{delf}).
It contradicts to the basic supposition of the global defect 
that the perturbation $\delta f$
becomes small as the radius $r$ increases at asymptotic region and reaches
zero at spatial infinity.
Therefore, the static regular global U(1) vortex cannot be compatible with
the vacuum brane world of an exponentially decaying warp factor, that is 
consistent with Ref.~\cite{Gregory:1999gv}. 
Similarly, when both $A^{'}(\infty)$ and $C(\infty)$ are nonvanishing
constants as in Eqs.~(\ref{aGre})--(\ref{cGre}), Eq.~(\ref{delf}) produces 
contradiction as follows:
\begin{eqnarray}
\delta f^{''} \pm \frac{2(p+1)\omega}{\sqrt{p+2}} \delta f^{'} - \left[
\frac{\omega^2}{2 \pi G_D (p+2)} +2 \right] \delta f &\approx& -
\frac{\omega^2}{2 \pi G_D (p+2)}
\nonumber\\
\stackrel{r\rightarrow\infty}{\Longrightarrow}~~ 
0&=&- \frac{|\Lambda|}{4 \pi G_D (p+1)}, 
\end{eqnarray}
where $\omega$ is replaced by Eq.~(\ref{omega}) in the last step of 
taking infinite radius. It means
that the cigar-like brane world predicted in Ref.~\cite{Gregory:1999gv} 
may not be supported by a global vortex but it is consistent with 
Ref.~\cite{Cohen:1999ia} in
the limit of vanishing cosmological constant, which involves a physical
singularity believed to be harmless.
In synthesis, we conclude that we could not find any regular static global
U(1) vortex solution to generate a proper brane world of an
exponentially decaying warp factor, yet, as far as a negative cosmological 
constant is turned on. 

If we remind of the expansion of the scalar field and the metric functions
near the origin (\ref{f0r})--(\ref{c0r}), which cannot be consistent with the
geometry of an exponentially decaying warp factor but can be compatible with
that of an exponentially growing warp factor, it may be intriguing to look
into a global vortex of which asymptote is the geometry of an
exponentially decaying warp factor. From such vacuum solutions, $A_{+}$ and
$C_{+}$ in Eqs.~(\ref{pasol})--(\ref{pcsol}) and $A$ and $C$ in
Eqs.~(\ref{areq})--(\ref{creq}), their leading asymptotic behavior is 
summarized as 
$A^{'}(r)\sim 2\omega/(p+2)$ and $C(r)\sim c_{\infty}\exp[-2\omega r/(p+2)]$.
Inserting these into Eq.~(\ref{delf}), we have
\begin{equation}
\delta f^{''} + 2\omega \delta f^{'} -2 \delta f \approx 0.
\end{equation}
This linear equation excludes power law asymptotic behavior of the scalar 
field, $\delta f\sim~\!\!\!\!\!\!\!/\;{\cal O}(1/r^{m})$, but it includes
an exponentially decaying asymptote, $\delta f\sim
\exp[-(\omega+\sqrt{\omega^{2}+2})r]$, which is different from
power law decaying of $\delta f$ for the global vortices in both flat and 
curved spacetime without a cosmological constant. An unexpected exponential 
approach of the scalar amplitude is allowable in this coordinate system with a 
warp factor due to  the exponential increasing of the radial 
circumference $C(r)$ at the asymptotic
region, i.e., the topological term proportional to $n^{2}/C^{2}$ in the scalar 
equation (\ref{scaleq}) goes to zero
in an exponential form for sufficiently-large $r$. Since this possibility 
satisfies boundary behaviors near both the 
origin and spatial infinity, there may exist
a regular configuration to connect these two boundaries. Since $n/c_{0}$
should be a natural number in the expansion (\ref{f0r})--(\ref{c0r}) near the
origin, $c_{0}$ should be chosen. There remains only one free parameter
$f_{0}$ in addition to three boundary values of $f(0)$, $A^{'}(0)$, $C(0)$,
while we need five input parameters to obtain a solution from two second-order
equations, (\ref{scaleq})--(\ref{simeq3}), and one first-order equation,
(\ref{simeq2}). The only adjustable parameter in Eqs.~(\ref{f0r})--(\ref{c0r})
is $\tilde{\omega}$ in Eq.~(\ref{omt}), which is expressed by a difference of
two fundamental scales, the $D$-dimensional Newton's constant 
$G_{D}$ and a cosmological constant
$\Lambda$. If there exists such a solution, then it is obtainable by a fine
tuning of one of the two fundamental scales.
It means that we usually obtain a solution with positive $A(r)$ for large $r$, 
which is connected to the pure AdS geometry with an 
exponentially growing warp factor. 

Though the metric $C(r)$ increases
exponentially, but ratios, $C^{'}/C$ and $C^{''}/C$, are finite and smooth
at entire range.
Therefore, Kretschmann invariant is finite everywhere:
\begin{eqnarray}\label{sKret}
R^{ABCD}R_{ABCD} =  4\left[ 2{A^{'}}^2
\left(5{A^{'}}^2  + 4 {A^{''}} + 2 \frac{{C^{'}}^2}{C^2}\right)
+ \frac{{C^{''}}^2}{C^2} 
\right],
\end{eqnarray}
and thereby all the spacetime points are not physically singular. 

{}From the previous argument we have arrived at a no-go theorem that
any regular static bulk global U(1) vortex cannot form a brane world of
an exponentially decaying warp factor as far as the boundary condition 
of the scalar amplitude, $f(\infty)=1$, is kept. 
One possible exit is to obtain a regular static bulk global U(1) vortex 
solution to support an appropriate brane world by loosening boundary 
conditions as minimal as we can. Boundary value of the scalar
amplitude $f(\infty)=f_{\infty}$ is supposed to be at the range
$0<f_{\infty}<1$ because all possible spacetime boundary conditions for
the brane world geometry have not been compatible with $f(\infty)=1$.
Since asymptotic behavior of the scalar field $f$ is described by a small
perturbation $\delta f$ given as $f(r)\approx f_{\infty}-\delta f(r)$, the
scalar equation (\ref{scaleq}) is approximated as
\begin{eqnarray}\label{dfeq}
\delta f^{''}+\left[\frac{C^{'}}{C}+(p+1)A^{'}\right]\delta f^{'}
-\left[\frac{n^{2}}{C^{2}}+(3f^{2}_{\infty}-1)\right]\delta f\approx
-\left[\frac{n^{2}}{C^{2}}+(f^{2}_{\infty}-1)\right].
\end{eqnarray}
In the limit of infinite radius the coefficient of $\delta f^{'}$ term given by
derivatives of the metric functions is known to be finite, and thereby the
left-hand side of Eq.~(\ref{dfeq}) vanishes as $r\rightarrow \infty$. From the
right-hand side of Eq.~(\ref{dfeq}), we read 
\begin{eqnarray}\label{fin}
f_{\infty}=\sqrt{1-\frac{n^{2}}{C(\infty)^{2}}}.
\end{eqnarray}
If the metric function $C(r)$ approaches zero at spatial infinity as the
vacuum geometry of an exponentially decaying warp factor $C_{-}$ in
Eq.~(\ref{pcsol}), $f_{\infty}$ cannot be a real number 
so that the change of the scalar
amplitude at spatial infinity is not enough to produce a warp geometry
of the AdS vacuum. On the other hand, if $C(r)$ diverges at spatial
infinity, then $f_{\infty}=1$ which reduces to the original contradiction.
The only choice is to let $C(\infty)$ finite but larger than the vorticity $n$.

Though the boundary condition of the scalar field (\ref{fin}) 
is not familiar to us and we have
no clear physicswise motivation to accept such awkward soliton configuration 
yet, we try first to confirm the existence of such configuration.
A possible power series solution is assumed as follows:
\begin{eqnarray}
f(r)&\approx&f_{\infty}-\frac{f_{u}}{r^{u}}+\cdots
,\label{fsol}\\
A^{'}(r)&\approx&a_{\infty}+\frac{a_{v}}{r^{v}}+\cdots
,\label{asol}\\
C(r)&\approx&c_{\infty}+\frac{c_{w}}{r^{w}}+\cdots
,\label{csol}
\end{eqnarray}
where boundary values of the scalar field $f_{\infty}$ and the metric functions 
$a_{\infty},~c_{\infty}$ do not vanish. Since only
algebraic terms in the equations (\ref{scaleq}), (\ref{simeq2}),
(\ref{simeq3}) are needed for determining the leading terms so that a possible
set of boundary conditions at spatial infinity are given as
\begin{eqnarray}\label{finalb}
f_{\infty}=\sqrt{1-\xi^{2}}<1,~~a_{\infty}=-\frac{2}{p+1}
\sqrt{\frac{p|\Lambda|}{2p+1}},~~c_{\infty}=\frac{n}{\xi},
\end{eqnarray}
where $\xi=[|\Lambda|/2\pi(2p+1)G_{D}]^{1/4}$ and 
$2\pi(p+1)G_{D}>|\Lambda|$.
Since $C(\infty)=c_{\infty}$ is proportional to $n$, there is no constraint 
to the vorticity $n$ from the boundary condition (\ref{fin}).
From the leading behavior at asymptotic region, a cigar-like geometry 
after a rescaling of $x^{\mu}$ and $ds$ is summarized by a metric with an 
exponentially decaying warp factor:
\begin{eqnarray}\label{waco}
ds^{2}=\exp\left(-\frac{4}{p+1}\sqrt{\frac{p|\Lambda|}{2p+1}}\, r\right)
dx^{\mu}dx_{\mu}-dr^{2}
-n^{2}\sqrt{\frac{2\pi(2p+1)G_{D}}{|\Lambda|}}d\theta^{2}.
\end{eqnarray}

Systematic expansion up to the next order allows a unique series 
solution:
\begin{eqnarray}
f(r) &\approx& \sqrt{1-\xi^{2}}\left(1 + \frac{p}{5p+2} 
\frac{\xi^2 \eta}{1 - \xi^2}\frac{1}{r}+\cdots \right),
\label{ffso}\\
A^{'}(r) &\approx& -\frac{2}{p+1}\sqrt{\frac{p|\Lambda|}{2p+1}} \left(
1 + \frac{\eta}{5p+2} \frac{1}{r}+\cdots \right), \\
C(r) &\approx& \frac{n}{\xi}\left( 1 + \frac{p\eta}{5p+2}\frac{1}{r}+\cdots
\right),
\label{ccso}
\end{eqnarray}
where 
\begin{equation}
\eta = \frac{p+1}{2} \sqrt{\frac{2p+1}{p|\Lambda|}} \left[ 1+
\frac{4 p|\Lambda|}{(p+1)(2p+1)(1- \xi^2)} \right].
\end{equation}
Note that all the expansion parameters in Eqs.~(\ref{fsol})--(\ref{csol}) are 
determined by the Newton's constant $G_{D}$, a cosmological constant
$\Lambda$, and the vorticity $n$. It means that such configuration is achieved
by the aforementioned fine tuning. This phenomenon was already expected 
from the behavior of the expansion near the origin (\ref{f0r})--(\ref{c0r}).
In the next section, a clear explanation on this fine tuning will be given
after a coordinate transformation.

If we extend the form of metric functions to $A(r)\propto r^{s}$
$(s>1)$ and $C(r)\propto\exp\left(\tilde{c}r^{t_{1}}\right)$ or $C(r)\propto
r^{t_{2}}$, we encounter easily a mismatch in the expansion of series
solution. Furthermore, change of the functional behavior $f(r)$ from a
power series to an exponential form $f(r)\sim f_{\infty}-f_{e}\exp 
(-\hat{c}r)$
does not help for the generation of smooth solution anymore. Therefore, the
obtained solution (\ref{ffso})--(\ref{ccso}) 
must be unique regular static global U(1) vortex solution 
with a brane world geometry of an exponentially decaying warp factor up to now
by massaging the boundary condition (\ref{fin}) 
at a physicswisely-acceptable and minimal extent.

\setcounter{equation}{0}
\section{Black Brane World and Short Scalar Hair}

For the global $p$-brane obtained in the previous section, the scalar
amplitude $f$ did not reach the vacuum expectation value even at 
infinite $r$
as in Eq.~(\ref{finalb}). It means that there is a contribution from the 
scalar potential to nonvanishing effective cosmological constant at 
$r=\infty$
\begin{equation}
\lambda v^{2}\left[\Lambda +2\pi G_{D}({f_{\infty}}^{2}-1)^{2}\right]
=-\frac{2p}{1+2p}\lambda v^{2}|\Lambda|.
\end{equation}
In addition, boundary values of the metric functions (\ref{finalb}) are 
also different
from Eqs.~(\ref{aGre})--(\ref{cGre}).
Asymptotic geometry depicts qualitatively the same cigar-like (or 
cylindrical)
geometry and includes an exponentially decaying warp factor in front of the
$p$-brane coordinates except for a few quantitative changes: the radius of 
the cigar becomes $C(\infty)=n/[|\Lambda|/2\pi(2p+1)G_{D}]^{1/4}$ and 
the decaying in the exponential warp factor $A^{'}(\infty)=-\frac{2}{p+1}
\sqrt{\frac{p|\Lambda|}{2p+1}}$ at spatial infinity. 
Through the boundary condition of the scalar field at infinity of the 
extra-dimensions was drastically changed, the obtained asymptotic geometry is 
qualitatively the same as that under the assumption $f(\infty) = 
1$~\cite{Gregory:1999gv,Olasagasti:2000gx,
Gherghetta:2000jf,Moon:2000hn,Moon:2001ck,Gregory:2002tp}.

Now we have to address a question what kind of physical interpretation is
applicable to this awkward boundary condition of the scalar field $f$.
In the curved spacetime, a natural conjecture may be given as follows.
The bulk spacetime with a warp factor, 
obtained in this coordinate system (\ref{metric}), does not describe 
entire spacetime but a patch of it so that a point of spatial infinity in the 
extra-dimensions corresponds to a mid-point of the entire geometry which 
is describable in another coordinate system. 
An appropriate coordinate system may be the aforementioned Poincar\'{e} 
coordinates (\ref{metric1}), and, in this metric,
the equation of the scalar field $f$ (\ref{scaleq}) is written as
\begin{equation}\label{bscal}
B\frac{d^{2}f}{dR^{2}}+\left[\frac{B}{R}+\left(1+\frac{p}{2}\right)
\frac{dB}{dR}+(1+p)B\frac{d\Phi}{dR}\right]\frac{df}{dR}
-\left[\frac{n^{2}}{R^{2}}+(f^{2}-1)\right]f=0.
\end{equation}
As shown in Ref.~\cite{Moon:2000hn,Moon:2001ck,Kim:2004vt},
the metric functions involve a horizon at $R=R_{{\rm H}}$ and 
the near horizon geometry is shown to behave 
\begin{eqnarray}
B(R)\sim (R-R_{\rm H})^{2}~~\mbox{and}~~
\frac{d\Phi}{dR}\sim\frac{1}{R-R_{\rm H}}.
\end{eqnarray}
Therefore, presumably, the location of the horizon at $R_{{\rm H}}$ 
in the Poincar\'{e} coordinates (\ref{metric1}) corresponds to 
the infinity $(r\rightarrow \infty)$ of the metric with a warp factor
(\ref{metric}), and the scalar equation in both metrics, 
Eq.~(\ref{scaleq}) and Eq.~(\ref{bscal}),
reduce exactly to the following algebraic equations
at their boundaries:
\begin{equation}\label{algeq}
\left.\left[\frac{n^{2}}{C(r)^{2}}+(f(r)^{2}-1)\right]
\right|_{r\rightarrow\infty}=0
\Longleftrightarrow \left.\left[\frac{n^{2}}{R^{2}}+(f(R)^{2}-1)
\right]\right|_{R=R_{\rm H}}=0.
\end{equation}
Though our argument was based on a specific form of $\phi^{4}$ scalar 
potential, every argument can also be applied to any scalar potential 
$V(f)$ once Eq.~(\ref{algeq})
is replaced by
\begin{equation}
\left.\left[\frac{n^{2}}{C(r)^{2}}+\frac{1}{f}\frac{dV}{df}\right]
\right|_{r\rightarrow\infty}=0,
\end{equation}
and it contains a solution $f(\infty)$ smaller than the vacuum expectation
value.

It implies that the finite boundary value of $C$ at infinity should coincide
with the finite radius of the horizon $R_{\rm H}$ proportional to 
the vorticity $n$
\begin{eqnarray}
R_{{\rm H}}&=& \mathop{\lim} \limits_{r\to \infty}C(r)=\frac{n}{\xi},\\
f(R_{{\rm H}})&=&\mathop{\lim}\limits_{r\to \infty}f(r)=\sqrt{1-\xi^{2}} 
\, ,
\end{eqnarray}
where $\xi$ is given around Eq.~(\ref{finalb}).
Furthermore, at asymptotic region of $r$, performing a coordinate
transformation from Eq.~(\ref{waco})
\begin{equation}
e^{2a_{\infty}r}=(R_{{\rm H}}-\rho)^{2}
\end{equation}
with a rescaling, $d\tilde{s}=-a_{\infty}ds$ and $d\tilde{x}^{\mu}
=-a_{\infty}dx^{\mu}$, we obtain the metric of an extremal black $p$-brane
\begin{equation}
d\tilde{s}^{2}=(R_{{\rm H}}-\rho)^{2}d\tilde{x}^{\mu}d\tilde{x}_{\mu}
-\frac{d\rho^{2}}{(R_{{\rm H}}-\rho)^{2}}-R_{{\rm H}}^{2}
d(-a_{\infty}\theta)^{2},
\end{equation}
where $a_{\infty}$ is also given in Eq.~(\ref{finalb}). Since
$-a_{\infty}\ne 1$ in usual cases, a deficit angle is identified at the
generated horizon due to a long range topological term of scalar phase.
This phenomenon is consistent with the short scalar hair of extremal 
charged BTZ 
black hole~\cite{Kim:1997ba}, since it is nothing but a black $0$-brane in 
our context.

We give a rough explanation on the formation of extremal black $p$-brane
with degenerated horizon in the scheme of Hodge duality on the 
$p$-brane~\cite{Kaloper:2000xa} as has been done in $p=0$ 
case~\cite{Kim:1997ba}. In the $(p+3)$-dimensional bulk, the U(1) current 
$J_{A}$ is defined by and rewritten in terms of its dual $(p+2)$-form field
strength tensor $H^{A_{1}A_{2}\cdots A_{p+2}}$ as
\begin{equation}
J_{A}=v^{2}f^{2}\partial_{A}\Omega\sim \sqrt{-g}
\epsilon_{ABC_{0}C_{1}\cdots C_{p}}H^{BC_{0}C_{1}\cdots C_{p}}.
\end{equation}
Then, the corresponding Lagrange densities satisfy the following relation up to
a multiplicative constant
\begin{equation}\label{lden}
v^{2}f^{2}g^{AB}\partial_{A}\Omega\partial_{B}\Omega
\sim \frac{1}{v^{2}f^{2}}
g^{A_{1}B_{1}}g^{A_{2}B_{2}}\cdots g^{A_{p+2}B_{p+2}}
H_{A_{1}A_{2}\cdots A_{p+2}}H_{B_{1}B_{2}\cdots B_{p+2}}.
\end{equation}
The scalar phase $\Omega$ was given by $\Omega=n\theta$ for global defects
in the extra-dimensions, so the single electric component survives as 
follows $H_{t12\cdots pr}\sim C(r)^{-1}e^{(p+1)a_{\infty}r}v^{2}f^{2}n$. 
In the field strength tensor, $p$ transverse coordinates do not play any 
role except for introducing warp factors.
Substituting it into the Lagrange density (\ref{lden}), we see 
that at asymptotic region quadratic term of the dual $(p+2)$-form field 
strength tensor remains to be a constant term independent of the warp 
factor
\begin{eqnarray}
\frac{1}{v^{2}f^{2}}g^{rr}g^{tt}g^{11}\cdots g^{pp}H_{t12\cdots pr}^{2}
\sim v^{2}f^{2}g^{\theta\theta}(\partial_{\theta}\Omega)^{2}
=v^{2}f^{2}\frac{n^{2}}{C^{2}}\stackrel{r\rightarrow\infty}{\longrightarrow}
v^{2}f_{\infty}^{2}\frac{n^{2}}{c_{\infty}^{2}},
\end{eqnarray} 
and then structure of the Hodge duality is almost the same as that of any 
$p$.
At the origin of $r=0$, $1/f^{2}$-factor in Eq.~(\ref{lden}) plays an 
important role to make the site of $p$-brane regular. So there is a 
significant difference between the Hodge-dual field strength with the 
scalar amplitude $f$ and a pure $(p+2)$-form field strength
tensor since we should live on that singular $p$-brane in the latter case.

Let us examine the behavior of the scalar field and the metric functions near
the origin and show that those in both coordinate systems, 
Eq.~(\ref{metric}) and Eq.~(\ref{metric1}), behave consistently each other 
up to the leading
approximation. For the scalar field, it should vanish at the origin such that 
$f(r=0)=f(R=0)=0$ and its slope near the origin is the same as shown in
Eq.~(\ref{f0r}) and Ref.~\cite{Moon:2000hn}. Substituting the expansion 
of the metric
functions (\ref{aexp0})--(\ref{c0r}) into Eq.~(\ref{metric}), we obtain
\begin{equation}\label{r0me}
ds^{2}\approx \left(1-\frac{2\pi G_{D}-|\Lambda|}{p+1}r^{2}\right)
dx^{\mu}dx_{\mu} -dr^{2}-r^{2}d\theta^{2}.
\end{equation}
A coordinate transformation from Eq.~(\ref{r0me}) to the Poincar\'{e} 
coordinate system (\ref{metric1}) results in the same metric obtained by the
power series expansion in the Poincar\'{e} coordinate system.
Since all the field and the metric functions are smooth between the origin and
spatial infinity where the boundary values coincide, it leads to a 
conclusion that the obtained cigar-like bulk geometry with an 
exponentially decaying warp factor parametrized by $r$-coordinate ($0\le 
r\le \infty$) coincides in fact with a patch in the coordinate system of 
$R$ $(0\le R\le R_{\rm H})$,
which is interior region bounded by the extremal black hole horizon as 
was the case of the vacuum solution in the previous section.

A few comments on comparison of the geometric properties of two bulk 
spacetimes are in order. First,
infinite radial distance $\int_{0}^{\infty}dr$ in this brane world corresponds
to infinite proper distance to the horizon of the extremal black hole,
$\int_{0}^{R_{\rm H}}dR/\sqrt{B(R)}\sim\mathop{\lim}\limits_{R\to R_{{\rm H}}}
\ln(R_{{\rm H}}-R)\sim\infty$. 
Second, finiteness of the bulk volume of the brane
world divided by the spacetime volume of the $p$-brane such as
\begin{equation}\label{acvo}
\frac{\int d^{p+3}x\sqrt{|g_{p+3}|}}{\int d^{p+1}x}=2\pi\int_{0}^{\infty}dr
\, e^{(p+1)A(r)}C(r)\sim\mbox{finite}
\end{equation}
can be interpreted as the finite volume inside the horizon divided by the
spacetime volume of the $p$-brane:
\begin{equation}\label{bpvo}
\frac{\int d^{p+3}x\sqrt{|g_{p+3}|}}{\int d^{p+1}x}=2\pi\int_{0}^{R_{\rm H}}
dR\,Re^{(p+1)\Phi}B^{p/2}\sim\mbox{finite}.
\end{equation}
Since $A(r\rightarrow \infty)$ is independent of the vorticity $n$ and
$C(r\rightarrow \infty)$ is proportional to $n$ as given in Eq.~(\ref{finalb}),
the bulk volume of the brane
world divided by the spacetime volume of the $p$-brane
(\ref{acvo}) (or equivalently, Eq.~(\ref{bpvo})) can be very small
when $n$ is small and very large when $n$ becomes a huge number.
Noticing that the only variable parameter in 
the algebraic equation (\ref{algeq}) is topological charge $n$ 
for a given scale, the very process may provide a mechanism to 
interpolate the RS model I 
of small extra-dimensions to the ADD 
model of large extra-dimensions by quantity of 
the topological charge $n$. However, we have no physical reason to fix
the vorticity $n$ in our present scheme, so choice of an $n$ is also a 
fine-tuning though it seems mild.
 
The property that the scalar field can approach the vacuum value $f=1$
at spatial infinity is available only when the circumference proportional
to $C/n$ goes 
to infinity as has been read in Eq.~(\ref{algeq}). Since
$C(\infty)/n=1/[|\Lambda|/2\pi (2p+1)G_{D}]^{1/4}$, such a limit can be
achieved by the limit of vanishing $|\Lambda|/G_{D}$.
After a rescaling back to the original variables by using Eq.~(\ref{resc}),
we have $|\Lambda|/G_{D}\rightarrow \lambda|\Lambda|M_{*}^{p+1}$ which 
implies either $\lambda\rightarrow 0$ or $|\Lambda|M_{*}^{p+1}\rightarrow 0$
in order to have $\mathop{\lim}\limits_{r\rightarrow \infty}f(r)=1$.
Here $\lambda\rightarrow 0$ seems an unwanted situation. 
It means that the extremal black hole becomes free from the short hair of
scalar amplitude, but it seems not so interesting in the context
of black hole physics since the radius of the horizon moves far away 
to the infinity, $R_{\rm H}\rightarrow\infty$, by the right-hand side 
of Eq.~(\ref{algeq}). Furthermore,   
the limit of vanishing cosmological constant heads for appearance
a physical singularity at infinity~\cite{Cohen:1999ia}.
Let us emphasize again that the limit of 
$\mathop{\lim}\limits_{r\rightarrow \infty}f=1$
cannot be taken as far as $|\Lambda|/G_{D}$ is finite even if the 
topological charge
$n$ diverges, and thereby the existence of a short scalar hair is
unavoidable in the case of global U(1) defect without an unnatural 
fine tuning.

In relation with an extension to arbitrary $N(\ge 3)$ extra-dimensions with
O($N$) linear $\sigma$-model and their higher-dimensional
analogues~\cite{Olasagasti:2000gx,Moon:2000hn,Gherghetta:2000jf,Benson:2001ac}
probably develop a short scalar hair, since the topological term 
proportional to $n^2/C^2$ both in Eq.~(\ref{scaleq}) and Eq.~(\ref{simeq3}) 
exists even in higher extra-dimensions with unit topological charge 
$(n=1)$. The previous analysis of the
black brane world based on half $\sigma$-lumps in O($N+1$) nonlinear
$\sigma$-model~\cite{Kim:1998gx,Kim:2004vt} provides a positive hint on
a short scalar hair, of which results were independent of 
the extra-dimensions $N$.

\section{Discussion}

In this paper we considered a bulk complex scalar field in
a warp geometry and the global vortices in rotationally symmetric 
two extra-dimensions, which form a flat thick $p$-brane. The corresponding
brane world is identified with the interior of an extremal black brane world
bounded by a horizon. An unavoidable short hair of the scalar amplitude 
on the horizon is translated as its boundary value at infinity 
in the metric with an exponentially decaying warp factor. Despite this short 
scalar hair, the obtained warp geometry is qualitatively the same as that in 
Ref.~\cite{Gregory:1999gv}.

In section 2, we obtained general static brane world solutions with
two extra-dimensions, but it is intriguing to extend our results
including general time-dependent solutions as in the case of one 
extra-dimension~\cite{Kaloper:1999sm} or that in arbitrary extra 
dimensions~\cite{Ito:2001fd,Cho:2003en}.
These questions are also applied to 
the black $p$-brane world solutions formed by global vortices in the two
extra-dimensions. Particularly, fate of the short scalar hair should be asked
when time dependence is taken into account~\cite{Gregory:2002tp}.

In our approach we have assumed a bulk complex scalar field in 
(1+5)-dimensions, and considered a global vortex solution in two 
extra-dimensions to interpret our world as 3-brane of codimension-two.
Then gapless Goldstone modes can be generated
outside the horizon in the broken phase and, 
despite of the existence of this horizon, 
those energetic modes can reach the 3-brane of our world,
where they may threaten conservation of energy. 
This implies a limitation of this toy model based on local
field theory, and
it should be improved by the model from string theory.
A natural way is to understand our (1+5)-dimensional bulk with a complex 
scalar field as a space-filling D5-${\bar {\rm D}}$5 pair. The decent relations
among D-branes in string theory dictate the decay of D5-${\bar {\rm D}}$5 pair
to codimension-two brane and its dynamics is described by condensation of
a complex tachyon~\cite{Sen:1999mg}. After the D5-${\bar {\rm D}}$5 pair decays
into codimension-two brane, all the open string modes
including Goldstone modes should disappear in the extra-dimensions.
In such sense, an appropriate effective field theory model derived 
from string theory is chosen and should be tackled. Though the effective 
field theory of tachyon is usually depicted by Born-Infeld type nonlocal
action with runaway 
potential~\cite{Sen:2002an}, tachyon vortices may share similar
characters with global vortices in various features.

\section*{Acknowledgements}
This work is supported by Korea Research Foundation Grants
KRF-2001-015-DP0082 and
is the result of research activities (Astrophysical Research
Center for the Structure and Evolution of the Cosmos (ARCSEC))
supported by Korea Science $\&$ Engineering Foundation.

\end{document}

%% file: 6dfig1.tex
\begingroup%
  \makeatletter%
  \newcommand{\GNUPLOTspecial}{%
    \@sanitize\catcode`\%=14\relax\special}%
  \setlength{\unitlength}{0.1bp}%
{\GNUPLOTspecial{!
/gnudict 256 dict def
gnudict begin
/Color false def
/Solid false def
/gnulinewidth 5.000 def
/userlinewidth gnulinewidth def
/vshift -33 def
/dl {10 mul} def
/hpt_ 31.5 def
/vpt_ 31.5 def
/hpt hpt_ def
/vpt vpt_ def
/M {moveto} bind def
/L {lineto} bind def
/R {rmoveto} bind def
/V {rlineto} bind def
/vpt2 vpt 2 mul def
/hpt2 hpt 2 mul def
/Lshow { currentpoint stroke M
  0 vshift R show } def
/Rshow { currentpoint stroke M
  dup stringwidth pop neg vshift R show } def
/Cshow { currentpoint stroke M
  dup stringwidth pop -2 div vshift R show } def
/UP { dup vpt_ mul /vpt exch def hpt_ mul /hpt exch def
  /hpt2 hpt 2 mul def /vpt2 vpt 2 mul def } def
/DL { Color {setrgbcolor Solid {pop []} if 0 setdash }
 {pop pop pop Solid {pop []} if 0 setdash} ifelse } def
/BL { stroke userlinewidth 2 mul setlinewidth } def
/AL { stroke userlinewidth 2 div setlinewidth } def
/UL { dup gnulinewidth mul /userlinewidth exch def
      10 mul /udl exch def } def
/PL { stroke userlinewidth setlinewidth } def
/LTb { BL [] 0 0 0 DL } def
/LTa { AL [1 udl mul 2 udl mul] 0 setdash 0 0 0 setrgbcolor } def
/LT0 { PL [] 1 0 0 DL } def
/LT1 { PL [4 dl 2 dl] 0 1 0 DL } def
/LT2 { PL [2 dl 3 dl] 0 0 1 DL } def
/LT3 { PL [1 dl 1.5 dl] 1 0 1 DL } def
/LT4 { PL [5 dl 2 dl 1 dl 2 dl] 0 1 1 DL } def
/LT5 { PL [4 dl 3 dl 1 dl 3 dl] 1 1 0 DL } def
/LT6 { PL [2 dl 2 dl 2 dl 4 dl] 0 0 0 DL } def
/LT7 { PL [2 dl 2 dl 2 dl 2 dl 2 dl 4 dl] 1 0.3 0 DL } def
/LT8 { PL [2 dl 2 dl 2 dl 2 dl 2 dl 2 dl 2 dl 4 dl] 0.5 0.5 0.5 DL } def
/Pnt { stroke [] 0 setdash
   gsave 1 setlinecap M 0 0 V stroke grestore } def
/Dia { stroke [] 0 setdash 2 copy vpt add M
  hpt neg vpt neg V hpt vpt neg V
  hpt vpt V hpt neg vpt V closepath stroke
  Pnt } def
/Pls { stroke [] 0 setdash vpt sub M 0 vpt2 V
  currentpoint stroke M
  hpt neg vpt neg R hpt2 0 V stroke
  } def
/Box { stroke [] 0 setdash 2 copy exch hpt sub exch vpt add M
  0 vpt2 neg V hpt2 0 V 0 vpt2 V
  hpt2 neg 0 V closepath stroke
  Pnt } def
/Crs { stroke [] 0 setdash exch hpt sub exch vpt add M
  hpt2 vpt2 neg V currentpoint stroke M
  hpt2 neg 0 R hpt2 vpt2 V stroke } def
/TriU { stroke [] 0 setdash 2 copy vpt 1.12 mul add M
  hpt neg vpt -1.62 mul V
  hpt 2 mul 0 V
  hpt neg vpt 1.62 mul V closepath stroke
  Pnt  } def
/Star { 2 copy Pls Crs } def
/BoxF { stroke [] 0 setdash exch hpt sub exch vpt add M
  0 vpt2 neg V  hpt2 0 V  0 vpt2 V
  hpt2 neg 0 V  closepath fill } def
/TriUF { stroke [] 0 setdash vpt 1.12 mul add M
  hpt neg vpt -1.62 mul V
  hpt 2 mul 0 V
  hpt neg vpt 1.62 mul V closepath fill } def
/TriD { stroke [] 0 setdash 2 copy vpt 1.12 mul sub M
  hpt neg vpt 1.62 mul V
  hpt 2 mul 0 V
  hpt neg vpt -1.62 mul V closepath stroke
  Pnt  } def
/TriDF { stroke [] 0 setdash vpt 1.12 mul sub M
  hpt neg vpt 1.62 mul V
  hpt 2 mul 0 V
  hpt neg vpt -1.62 mul V closepath fill} def
/DiaF { stroke [] 0 setdash vpt add M
  hpt neg vpt neg V hpt vpt neg V
  hpt vpt V hpt neg vpt V closepath fill } def
/Pent { stroke [] 0 setdash 2 copy gsave
  translate 0 hpt M 4 {72 rotate 0 hpt L} repeat
  closepath stroke grestore Pnt } def
/PentF { stroke [] 0 setdash gsave
  translate 0 hpt M 4 {72 rotate 0 hpt L} repeat
  closepath fill grestore } def
/Circle { stroke [] 0 setdash 2 copy
  hpt 0 360 arc stroke Pnt } def
/CircleF { stroke [] 0 setdash hpt 0 360 arc fill } def
/C0 { BL [] 0 setdash 2 copy moveto vpt 90 450  arc } bind def
/C1 { BL [] 0 setdash 2 copy        moveto
       2 copy  vpt 0 90 arc closepath fill
               vpt 0 360 arc closepath } bind def
/C2 { BL [] 0 setdash 2 copy moveto
       2 copy  vpt 90 180 arc closepath fill
               vpt 0 360 arc closepath } bind def
/C3 { BL [] 0 setdash 2 copy moveto
       2 copy  vpt 0 180 arc closepath fill
               vpt 0 360 arc closepath } bind def
/C4 { BL [] 0 setdash 2 copy moveto
       2 copy  vpt 180 270 arc closepath fill
               vpt 0 360 arc closepath } bind def
/C5 { BL [] 0 setdash 2 copy moveto
       2 copy  vpt 0 90 arc
       2 copy moveto
       2 copy  vpt 180 270 arc closepath fill
               vpt 0 360 arc } bind def
/C6 { BL [] 0 setdash 2 copy moveto
      2 copy  vpt 90 270 arc closepath fill
              vpt 0 360 arc closepath } bind def
/C7 { BL [] 0 setdash 2 copy moveto
      2 copy  vpt 0 270 arc closepath fill
              vpt 0 360 arc closepath } bind def
/C8 { BL [] 0 setdash 2 copy moveto
      2 copy vpt 270 360 arc closepath fill
              vpt 0 360 arc closepath } bind def
/C9 { BL [] 0 setdash 2 copy moveto
      2 copy  vpt 270 450 arc closepath fill
              vpt 0 360 arc closepath } bind def
/C10 { BL [] 0 setdash 2 copy 2 copy moveto vpt 270 360 arc closepath fill
       2 copy moveto
       2 copy vpt 90 180 arc closepath fill
               vpt 0 360 arc closepath } bind def
/C11 { BL [] 0 setdash 2 copy moveto
       2 copy  vpt 0 180 arc closepath fill
       2 copy moveto
       2 copy  vpt 270 360 arc closepath fill
               vpt 0 360 arc closepath } bind def
/C12 { BL [] 0 setdash 2 copy moveto
       2 copy  vpt 180 360 arc closepath fill
               vpt 0 360 arc closepath } bind def
/C13 { BL [] 0 setdash  2 copy moveto
       2 copy  vpt 0 90 arc closepath fill
       2 copy moveto
       2 copy  vpt 180 360 arc closepath fill
               vpt 0 360 arc closepath } bind def
/C14 { BL [] 0 setdash 2 copy moveto
       2 copy  vpt 90 360 arc closepath fill
               vpt 0 360 arc } bind def
/C15 { BL [] 0 setdash 2 copy vpt 0 360 arc closepath fill
               vpt 0 360 arc closepath } bind def
/Rec   { newpath 4 2 roll moveto 1 index 0 rlineto 0 exch rlineto
       neg 0 rlineto closepath } bind def
/Square { dup Rec } bind def
/Bsquare { vpt sub exch vpt sub exch vpt2 Square } bind def
/S0 { BL [] 0 setdash 2 copy moveto 0 vpt rlineto BL Bsquare } bind def
/S1 { BL [] 0 setdash 2 copy vpt Square fill Bsquare } bind def
/S2 { BL [] 0 setdash 2 copy exch vpt sub exch vpt Square fill Bsquare } bind def
/S3 { BL [] 0 setdash 2 copy exch vpt sub exch vpt2 vpt Rec fill Bsquare } bind def
/S4 { BL [] 0 setdash 2 copy exch vpt sub exch vpt sub vpt Square fill Bsquare } bind def
/S5 { BL [] 0 setdash 2 copy 2 copy vpt Square fill
       exch vpt sub exch vpt sub vpt Square fill Bsquare } bind def
/S6 { BL [] 0 setdash 2 copy exch vpt sub exch vpt sub vpt vpt2 Rec fill Bsquare } bind def
/S7 { BL [] 0 setdash 2 copy exch vpt sub exch vpt sub vpt vpt2 Rec fill
       2 copy vpt Square fill
       Bsquare } bind def
/S8 { BL [] 0 setdash 2 copy vpt sub vpt Square fill Bsquare } bind def
/S9 { BL [] 0 setdash 2 copy vpt sub vpt vpt2 Rec fill Bsquare } bind def
/S10 { BL [] 0 setdash 2 copy vpt sub vpt Square fill 2 copy exch vpt sub exch vpt Square fill
       Bsquare } bind def
/S11 { BL [] 0 setdash 2 copy vpt sub vpt Square fill 2 copy exch vpt sub exch vpt2 vpt Rec fill
       Bsquare } bind def
/S12 { BL [] 0 setdash 2 copy exch vpt sub exch vpt sub vpt2 vpt Rec fill Bsquare } bind def
/S13 { BL [] 0 setdash 2 copy exch vpt sub exch vpt sub vpt2 vpt Rec fill
       2 copy vpt Square fill Bsquare } bind def
/S14 { BL [] 0 setdash 2 copy exch vpt sub exch vpt sub vpt2 vpt Rec fill
       2 copy exch vpt sub exch vpt Square fill Bsquare } bind def
/S15 { BL [] 0 setdash 2 copy Bsquare fill Bsquare } bind def
/D0 { gsave translate 45 rotate 0 0 S0 stroke grestore } bind def
/D1 { gsave translate 45 rotate 0 0 S1 stroke grestore } bind def
/D2 { gsave translate 45 rotate 0 0 S2 stroke grestore } bind def
/D3 { gsave translate 45 rotate 0 0 S3 stroke grestore } bind def
/D4 { gsave translate 45 rotate 0 0 S4 stroke grestore } bind def
/D5 { gsave translate 45 rotate 0 0 S5 stroke grestore } bind def
/D6 { gsave translate 45 rotate 0 0 S6 stroke grestore } bind def
/D7 { gsave translate 45 rotate 0 0 S7 stroke grestore } bind def
/D8 { gsave translate 45 rotate 0 0 S8 stroke grestore } bind def
/D9 { gsave translate 45 rotate 0 0 S9 stroke grestore } bind def
/D10 { gsave translate 45 rotate 0 0 S10 stroke grestore } bind def
/D11 { gsave translate 45 rotate 0 0 S11 stroke grestore } bind def
/D12 { gsave translate 45 rotate 0 0 S12 stroke grestore } bind def
/D13 { gsave translate 45 rotate 0 0 S13 stroke grestore } bind def
/D14 { gsave translate 45 rotate 0 0 S14 stroke grestore } bind def
/D15 { gsave translate 45 rotate 0 0 S15 stroke grestore } bind def
/DiaE { stroke [] 0 setdash vpt add M
  hpt neg vpt neg V hpt vpt neg V
  hpt vpt V hpt neg vpt V closepath stroke } def
/BoxE { stroke [] 0 setdash exch hpt sub exch vpt add M
  0 vpt2 neg V hpt2 0 V 0 vpt2 V
  hpt2 neg 0 V closepath stroke } def
/TriUE { stroke [] 0 setdash vpt 1.12 mul add M
  hpt neg vpt -1.62 mul V
  hpt 2 mul 0 V
  hpt neg vpt 1.62 mul V closepath stroke } def
/TriDE { stroke [] 0 setdash vpt 1.12 mul sub M
  hpt neg vpt 1.62 mul V
  hpt 2 mul 0 V
  hpt neg vpt -1.62 mul V closepath stroke } def
/PentE { stroke [] 0 setdash gsave
  translate 0 hpt M 4 {72 rotate 0 hpt L} repeat
  closepath stroke grestore } def
/CircE { stroke [] 0 setdash 
  hpt 0 360 arc stroke } def
/Opaque { gsave closepath 1 setgray fill grestore 0 setgray closepath } def
/DiaW { stroke [] 0 setdash vpt add M
  hpt neg vpt neg V hpt vpt neg V
  hpt vpt V hpt neg vpt V Opaque stroke } def
/BoxW { stroke [] 0 setdash exch hpt sub exch vpt add M
  0 vpt2 neg V hpt2 0 V 0 vpt2 V
  hpt2 neg 0 V Opaque stroke } def
/TriUW { stroke [] 0 setdash vpt 1.12 mul add M
  hpt neg vpt -1.62 mul V
  hpt 2 mul 0 V
  hpt neg vpt 1.62 mul V Opaque stroke } def
/TriDW { stroke [] 0 setdash vpt 1.12 mul sub M
  hpt neg vpt 1.62 mul V
  hpt 2 mul 0 V
  hpt neg vpt -1.62 mul V Opaque stroke } def
/PentW { stroke [] 0 setdash gsave
  translate 0 hpt M 4 {72 rotate 0 hpt L} repeat
  Opaque stroke grestore } def
/CircW { stroke [] 0 setdash 
  hpt 0 360 arc Opaque stroke } def
/BoxFill { gsave Rec 1 setgray fill grestore } def
end
}}%
\begin{picture}(3600,2160)(0,0)%
{\GNUPLOTspecial{"
gnudict begin
gsave
0 0 translate
0.100 0.100 scale
0 setgray
newpath
1.000 UL
LTb
1.000 UL
LT0
3175 100 M
550 0 V
3603 67 M
122 33 V
-122 32 V
1.000 UL
LT0
150 1815 M
0 490 V
32 -122 R
-32 122 V
117 2183 L
1.000 UL
LT0
150 590 M
33 -12 V
34 -12 V
33 -11 V
33 -11 V
34 -11 V
33 -11 V
33 -10 V
34 -10 V
33 -10 V
33 -10 V
34 -9 V
33 -9 V
33 -9 V
34 -9 V
33 -8 V
33 -9 V
34 -8 V
33 -8 V
33 -7 V
34 -8 V
33 -7 V
33 -7 V
34 -7 V
33 -7 V
33 -7 V
34 -6 V
33 -6 V
33 -7 V
34 -6 V
33 -5 V
33 -6 V
34 -6 V
33 -5 V
33 -5 V
34 -6 V
33 -5 V
33 -4 V
34 -5 V
33 -5 V
33 -5 V
34 -4 V
33 -4 V
33 -5 V
34 -4 V
33 -4 V
33 -4 V
34 -3 V
33 -4 V
33 -4 V
34 -3 V
33 -4 V
33 -3 V
34 -4 V
33 -3 V
33 -3 V
34 -3 V
33 -3 V
33 -3 V
34 -3 V
33 -3 V
33 -2 V
34 -3 V
33 -2 V
33 -3 V
34 -2 V
33 -3 V
33 -2 V
34 -2 V
33 -3 V
33 -2 V
34 -2 V
33 -2 V
33 -2 V
34 -2 V
33 -2 V
33 -2 V
34 -2 V
33 -1 V
33 -2 V
34 -2 V
33 -1 V
33 -2 V
34 -2 V
33 -1 V
33 -2 V
34 -1 V
33 -2 V
33 -1 V
34 -1 V
33 -2 V
33 -1 V
34 -1 V
33 -1 V
33 -2 V
34 -1 V
33 -1 V
33 -1 V
34 -1 V
33 -1 V
1.000 UL
LT1
150 590 M
33 12 V
34 13 V
33 13 V
33 13 V
34 14 V
33 14 V
33 14 V
34 15 V
33 15 V
33 15 V
34 16 V
33 16 V
33 17 V
34 17 V
33 17 V
33 18 V
34 18 V
33 19 V
33 19 V
34 20 V
33 20 V
33 21 V
34 21 V
33 22 V
33 23 V
34 23 V
33 23 V
33 24 V
34 25 V
33 25 V
33 26 V
34 27 V
33 27 V
33 28 V
34 29 V
33 29 V
33 30 V
34 31 V
33 32 V
33 32 V
34 33 V
33 34 V
33 35 V
34 36 V
33 37 V
33 38 V
34 38 V
33 40 V
33 40 V
34 42 V
33 43 V
33 43 V
34 45 V
33 46 V
33 47 V
28 40 V
1.000 UL
LT4
150 790 M
33 -12 V
34 -12 V
33 -12 V
33 -11 V
34 -11 V
33 -10 V
33 -10 V
34 -10 V
33 -10 V
33 -9 V
34 -8 V
33 -8 V
33 -8 V
34 -8 V
33 -7 V
33 -7 V
34 -6 V
33 -6 V
33 -5 V
34 -5 V
33 -5 V
33 -4 V
34 -4 V
33 -3 V
33 -2 V
34 -3 V
33 -2 V
33 -1 V
34 -1 V
33 0 V
33 0 V
34 1 V
33 1 V
33 1 V
34 3 V
33 2 V
33 3 V
34 4 V
33 3 V
33 5 V
34 5 V
33 5 V
33 6 V
34 6 V
33 6 V
33 7 V
34 8 V
33 7 V
33 8 V
34 9 V
33 9 V
33 9 V
34 10 V
33 10 V
33 10 V
34 11 V
33 11 V
33 11 V
34 12 V
33 12 V
33 12 V
34 13 V
33 13 V
33 14 V
34 14 V
33 14 V
33 14 V
34 15 V
33 15 V
33 16 V
34 16 V
33 16 V
33 17 V
34 17 V
33 17 V
33 18 V
34 19 V
33 18 V
33 19 V
34 20 V
33 19 V
33 21 V
34 20 V
33 21 V
33 22 V
34 22 V
33 22 V
33 23 V
34 24 V
33 24 V
33 24 V
34 25 V
33 25 V
33 26 V
34 26 V
33 27 V
33 28 V
34 28 V
33 28 V
1.000 UL
LT3
150 790 M
33 13 V
34 12 V
33 14 V
33 13 V
34 14 V
33 15 V
33 14 V
34 15 V
33 16 V
33 15 V
34 17 V
33 16 V
33 17 V
34 17 V
33 18 V
33 18 V
34 18 V
33 19 V
33 19 V
34 19 V
33 20 V
33 21 V
34 21 V
33 21 V
33 22 V
34 22 V
33 23 V
33 23 V
34 23 V
33 24 V
33 25 V
34 25 V
33 25 V
33 26 V
34 27 V
33 27 V
33 28 V
34 28 V
33 29 V
33 29 V
34 30 V
33 31 V
33 31 V
34 32 V
33 33 V
33 33 V
34 34 V
33 34 V
33 35 V
34 36 V
33 37 V
33 37 V
34 38 V
17 21 V
1.000 UL
LT0
150 100 M
3300 0 V
1.000 UL
LT0
150 2060 M
150 100 L
stroke
grestore
end
showpage
}}%
\put(3770,100){\makebox(0,0)[l]{\large $r$}}%
\put(2940,296){\makebox(0,0)[r]{\large $e^{2A_{-}(r)}$}}%
\put(1900,1423){\makebox(0,0)[r]{\large $e^{2A_{+}(r)}$}}%
\put(2650,1581){\makebox(0,0)[l]{\large $({\kappa}\ne0,\gamma>0)$}}%
\put(1635,1911){\makebox(0,0)[r]{\large $({\kappa}\ne0,\gamma<0)$}}%
\put(95,2158){\makebox(0,0)[r]{\large $e^{2A(r)}$}}%
\put(100,100){\makebox(0,0)[r]{\large $0$}}%
\end{picture}%
\endgroup
 

%% file: 6dfig2.tex
\begingroup%
  \makeatletter%
  \newcommand{\GNUPLOTspecial}{%
    \@sanitize\catcode`\%=14\relax\special}%
  \setlength{\unitlength}{0.1bp}%
{\GNUPLOTspecial{!
/gnudict 256 dict def
gnudict begin
/Color false def
/Solid false def
/gnulinewidth 5.000 def
/userlinewidth gnulinewidth def
/vshift -33 def
/dl {10 mul} def
/hpt_ 31.5 def
/vpt_ 31.5 def
/hpt hpt_ def
/vpt vpt_ def
/M {moveto} bind def
/L {lineto} bind def
/R {rmoveto} bind def
/V {rlineto} bind def
/vpt2 vpt 2 mul def
/hpt2 hpt 2 mul def
/Lshow { currentpoint stroke M
  0 vshift R show } def
/Rshow { currentpoint stroke M
  dup stringwidth pop neg vshift R show } def
/Cshow { currentpoint stroke M
  dup stringwidth pop -2 div vshift R show } def
/UP { dup vpt_ mul /vpt exch def hpt_ mul /hpt exch def
  /hpt2 hpt 2 mul def /vpt2 vpt 2 mul def } def
/DL { Color {setrgbcolor Solid {pop []} if 0 setdash }
 {pop pop pop Solid {pop []} if 0 setdash} ifelse } def
/BL { stroke userlinewidth 2 mul setlinewidth } def
/AL { stroke userlinewidth 2 div setlinewidth } def
/UL { dup gnulinewidth mul /userlinewidth exch def
      10 mul /udl exch def } def
/PL { stroke userlinewidth setlinewidth } def
/LTb { BL [] 0 0 0 DL } def
/LTa { AL [1 udl mul 2 udl mul] 0 setdash 0 0 0 setrgbcolor } def
/LT0 { PL [] 1 0 0 DL } def
/LT1 { PL [4 dl 2 dl] 0 1 0 DL } def
/LT2 { PL [2 dl 3 dl] 0 0 1 DL } def
/LT3 { PL [1 dl 1.5 dl] 1 0 1 DL } def
/LT4 { PL [5 dl 2 dl 1 dl 2 dl] 0 1 1 DL } def
/LT5 { PL [4 dl 3 dl 1 dl 3 dl] 1 1 0 DL } def
/LT6 { PL [2 dl 2 dl 2 dl 4 dl] 0 0 0 DL } def
/LT7 { PL [2 dl 2 dl 2 dl 2 dl 2 dl 4 dl] 1 0.3 0 DL } def
/LT8 { PL [2 dl 2 dl 2 dl 2 dl 2 dl 2 dl 2 dl 4 dl] 0.5 0.5 0.5 DL } def
/Pnt { stroke [] 0 setdash
   gsave 1 setlinecap M 0 0 V stroke grestore } def
/Dia { stroke [] 0 setdash 2 copy vpt add M
  hpt neg vpt neg V hpt vpt neg V
  hpt vpt V hpt neg vpt V closepath stroke
  Pnt } def
/Pls { stroke [] 0 setdash vpt sub M 0 vpt2 V
  currentpoint stroke M
  hpt neg vpt neg R hpt2 0 V stroke
  } def
/Box { stroke [] 0 setdash 2 copy exch hpt sub exch vpt add M
  0 vpt2 neg V hpt2 0 V 0 vpt2 V
  hpt2 neg 0 V closepath stroke
  Pnt } def
/Crs { stroke [] 0 setdash exch hpt sub exch vpt add M
  hpt2 vpt2 neg V currentpoint stroke M
  hpt2 neg 0 R hpt2 vpt2 V stroke } def
/TriU { stroke [] 0 setdash 2 copy vpt 1.12 mul add M
  hpt neg vpt -1.62 mul V
  hpt 2 mul 0 V
  hpt neg vpt 1.62 mul V closepath stroke
  Pnt  } def
/Star { 2 copy Pls Crs } def
/BoxF { stroke [] 0 setdash exch hpt sub exch vpt add M
  0 vpt2 neg V  hpt2 0 V  0 vpt2 V
  hpt2 neg 0 V  closepath fill } def
/TriUF { stroke [] 0 setdash vpt 1.12 mul add M
  hpt neg vpt -1.62 mul V
  hpt 2 mul 0 V
  hpt neg vpt 1.62 mul V closepath fill } def
/TriD { stroke [] 0 setdash 2 copy vpt 1.12 mul sub M
  hpt neg vpt 1.62 mul V
  hpt 2 mul 0 V
  hpt neg vpt -1.62 mul V closepath stroke
  Pnt  } def
/TriDF { stroke [] 0 setdash vpt 1.12 mul sub M
  hpt neg vpt 1.62 mul V
  hpt 2 mul 0 V
  hpt neg vpt -1.62 mul V closepath fill} def
/DiaF { stroke [] 0 setdash vpt add M
  hpt neg vpt neg V hpt vpt neg V
  hpt vpt V hpt neg vpt V closepath fill } def
/Pent { stroke [] 0 setdash 2 copy gsave
  translate 0 hpt M 4 {72 rotate 0 hpt L} repeat
  closepath stroke grestore Pnt } def
/PentF { stroke [] 0 setdash gsave
  translate 0 hpt M 4 {72 rotate 0 hpt L} repeat
  closepath fill grestore } def
/Circle { stroke [] 0 setdash 2 copy
  hpt 0 360 arc stroke Pnt } def
/CircleF { stroke [] 0 setdash hpt 0 360 arc fill } def
/C0 { BL [] 0 setdash 2 copy moveto vpt 90 450  arc } bind def
/C1 { BL [] 0 setdash 2 copy        moveto
       2 copy  vpt 0 90 arc closepath fill
               vpt 0 360 arc closepath } bind def
/C2 { BL [] 0 setdash 2 copy moveto
       2 copy  vpt 90 180 arc closepath fill
               vpt 0 360 arc closepath } bind def
/C3 { BL [] 0 setdash 2 copy moveto
       2 copy  vpt 0 180 arc closepath fill
               vpt 0 360 arc closepath } bind def
/C4 { BL [] 0 setdash 2 copy moveto
       2 copy  vpt 180 270 arc closepath fill
               vpt 0 360 arc closepath } bind def
/C5 { BL [] 0 setdash 2 copy moveto
       2 copy  vpt 0 90 arc
       2 copy moveto
       2 copy  vpt 180 270 arc closepath fill
               vpt 0 360 arc } bind def
/C6 { BL [] 0 setdash 2 copy moveto
      2 copy  vpt 90 270 arc closepath fill
              vpt 0 360 arc closepath } bind def
/C7 { BL [] 0 setdash 2 copy moveto
      2 copy  vpt 0 270 arc closepath fill
              vpt 0 360 arc closepath } bind def
/C8 { BL [] 0 setdash 2 copy moveto
      2 copy vpt 270 360 arc closepath fill
              vpt 0 360 arc closepath } bind def
/C9 { BL [] 0 setdash 2 copy moveto
      2 copy  vpt 270 450 arc closepath fill
              vpt 0 360 arc closepath } bind def
/C10 { BL [] 0 setdash 2 copy 2 copy moveto vpt 270 360 arc closepath fill
       2 copy moveto
       2 copy vpt 90 180 arc closepath fill
               vpt 0 360 arc closepath } bind def
/C11 { BL [] 0 setdash 2 copy moveto
       2 copy  vpt 0 180 arc closepath fill
       2 copy moveto
       2 copy  vpt 270 360 arc closepath fill
               vpt 0 360 arc closepath } bind def
/C12 { BL [] 0 setdash 2 copy moveto
       2 copy  vpt 180 360 arc closepath fill
               vpt 0 360 arc closepath } bind def
/C13 { BL [] 0 setdash  2 copy moveto
       2 copy  vpt 0 90 arc closepath fill
       2 copy moveto
       2 copy  vpt 180 360 arc closepath fill
               vpt 0 360 arc closepath } bind def
/C14 { BL [] 0 setdash 2 copy moveto
       2 copy  vpt 90 360 arc closepath fill
               vpt 0 360 arc } bind def
/C15 { BL [] 0 setdash 2 copy vpt 0 360 arc closepath fill
               vpt 0 360 arc closepath } bind def
/Rec   { newpath 4 2 roll moveto 1 index 0 rlineto 0 exch rlineto
       neg 0 rlineto closepath } bind def
/Square { dup Rec } bind def
/Bsquare { vpt sub exch vpt sub exch vpt2 Square } bind def
/S0 { BL [] 0 setdash 2 copy moveto 0 vpt rlineto BL Bsquare } bind def
/S1 { BL [] 0 setdash 2 copy vpt Square fill Bsquare } bind def
/S2 { BL [] 0 setdash 2 copy exch vpt sub exch vpt Square fill Bsquare } bind def
/S3 { BL [] 0 setdash 2 copy exch vpt sub exch vpt2 vpt Rec fill Bsquare } bind def
/S4 { BL [] 0 setdash 2 copy exch vpt sub exch vpt sub vpt Square fill Bsquare } bind def
/S5 { BL [] 0 setdash 2 copy 2 copy vpt Square fill
       exch vpt sub exch vpt sub vpt Square fill Bsquare } bind def
/S6 { BL [] 0 setdash 2 copy exch vpt sub exch vpt sub vpt vpt2 Rec fill Bsquare } bind def
/S7 { BL [] 0 setdash 2 copy exch vpt sub exch vpt sub vpt vpt2 Rec fill
       2 copy vpt Square fill
       Bsquare } bind def
/S8 { BL [] 0 setdash 2 copy vpt sub vpt Square fill Bsquare } bind def
/S9 { BL [] 0 setdash 2 copy vpt sub vpt vpt2 Rec fill Bsquare } bind def
/S10 { BL [] 0 setdash 2 copy vpt sub vpt Square fill 2 copy exch vpt sub exch vpt Square fill
       Bsquare } bind def
/S11 { BL [] 0 setdash 2 copy vpt sub vpt Square fill 2 copy exch vpt sub exch vpt2 vpt Rec fill
       Bsquare } bind def
/S12 { BL [] 0 setdash 2 copy exch vpt sub exch vpt sub vpt2 vpt Rec fill Bsquare } bind def
/S13 { BL [] 0 setdash 2 copy exch vpt sub exch vpt sub vpt2 vpt Rec fill
       2 copy vpt Square fill Bsquare } bind def
/S14 { BL [] 0 setdash 2 copy exch vpt sub exch vpt sub vpt2 vpt Rec fill
       2 copy exch vpt sub exch vpt Square fill Bsquare } bind def
/S15 { BL [] 0 setdash 2 copy Bsquare fill Bsquare } bind def
/D0 { gsave translate 45 rotate 0 0 S0 stroke grestore } bind def
/D1 { gsave translate 45 rotate 0 0 S1 stroke grestore } bind def
/D2 { gsave translate 45 rotate 0 0 S2 stroke grestore } bind def
/D3 { gsave translate 45 rotate 0 0 S3 stroke grestore } bind def
/D4 { gsave translate 45 rotate 0 0 S4 stroke grestore } bind def
/D5 { gsave translate 45 rotate 0 0 S5 stroke grestore } bind def
/D6 { gsave translate 45 rotate 0 0 S6 stroke grestore } bind def
/D7 { gsave translate 45 rotate 0 0 S7 stroke grestore } bind def
/D8 { gsave translate 45 rotate 0 0 S8 stroke grestore } bind def
/D9 { gsave translate 45 rotate 0 0 S9 stroke grestore } bind def
/D10 { gsave translate 45 rotate 0 0 S10 stroke grestore } bind def
/D11 { gsave translate 45 rotate 0 0 S11 stroke grestore } bind def
/D12 { gsave translate 45 rotate 0 0 S12 stroke grestore } bind def
/D13 { gsave translate 45 rotate 0 0 S13 stroke grestore } bind def
/D14 { gsave translate 45 rotate 0 0 S14 stroke grestore } bind def
/D15 { gsave translate 45 rotate 0 0 S15 stroke grestore } bind def
/DiaE { stroke [] 0 setdash vpt add M
  hpt neg vpt neg V hpt vpt neg V
  hpt vpt V hpt neg vpt V closepath stroke } def
/BoxE { stroke [] 0 setdash exch hpt sub exch vpt add M
  0 vpt2 neg V hpt2 0 V 0 vpt2 V
  hpt2 neg 0 V closepath stroke } def
/TriUE { stroke [] 0 setdash vpt 1.12 mul add M
  hpt neg vpt -1.62 mul V
  hpt 2 mul 0 V
  hpt neg vpt 1.62 mul V closepath stroke } def
/TriDE { stroke [] 0 setdash vpt 1.12 mul sub M
  hpt neg vpt 1.62 mul V
  hpt 2 mul 0 V
  hpt neg vpt -1.62 mul V closepath stroke } def
/PentE { stroke [] 0 setdash gsave
  translate 0 hpt M 4 {72 rotate 0 hpt L} repeat
  closepath stroke grestore } def
/CircE { stroke [] 0 setdash 
  hpt 0 360 arc stroke } def
/Opaque { gsave closepath 1 setgray fill grestore 0 setgray closepath } def
/DiaW { stroke [] 0 setdash vpt add M
  hpt neg vpt neg V hpt vpt neg V
  hpt vpt V hpt neg vpt V Opaque stroke } def
/BoxW { stroke [] 0 setdash exch hpt sub exch vpt add M
  0 vpt2 neg V hpt2 0 V 0 vpt2 V
  hpt2 neg 0 V Opaque stroke } def
/TriUW { stroke [] 0 setdash vpt 1.12 mul add M
  hpt neg vpt -1.62 mul V
  hpt 2 mul 0 V
  hpt neg vpt 1.62 mul V Opaque stroke } def
/TriDW { stroke [] 0 setdash vpt 1.12 mul sub M
  hpt neg vpt 1.62 mul V
  hpt 2 mul 0 V
  hpt neg vpt -1.62 mul V Opaque stroke } def
/PentW { stroke [] 0 setdash gsave
  translate 0 hpt M 4 {72 rotate 0 hpt L} repeat
  Opaque stroke grestore } def
/CircW { stroke [] 0 setdash 
  hpt 0 360 arc Opaque stroke } def
/BoxFill { gsave Rec 1 setgray fill grestore } def
end
}}%
\begin{picture}(3600,2160)(0,0)%
{\GNUPLOTspecial{"
gnudict begin
gsave
0 0 translate
0.100 0.100 scale
0 setgray
newpath
1.000 UL
LTb
1.000 UL
LT0
3450 100 M
165 0 V
3567 87 M
48 13 V
-48 12 V
1.000 UL
LT0
975 2060 M
0 168 V
13 -49 R
-13 49 V
-14 -49 V
1.000 UL
LT0
1140 688 M
975 380 L
23 102 R
975 380 L
71 76 V
1.000 UL
LT0
1792 436 M
-182 56 V
57 -3 R
-57 3 V
48 -31 V
1.000 UL
LT0
521 632 M
331 492 L
44 55 R
331 492 L
65 25 V
1.000 UL
LT0
1140 688 M
975 996 L
71 -77 R
-71 77 V
998 893 L
1.000 UL
LT0
521 632 M
331 646 L
56 10 R
331 646 L
53 -19 V
1.000 UL
LT0
1792 436 M
1610 646 L
69 -47 R
-69 47 V
36 -75 V
1.000 UL
LT4
333 496 M
22 -9 V
22 -9 V
23 -9 V
23 -8 V
23 -8 V
23 -7 V
24 -7 V
24 -6 V
25 -6 V
25 -6 V
25 -5 V
26 -5 V
26 -5 V
26 -4 V
26 -4 V
27 -3 V
27 -3 V
28 -3 V
27 -2 V
28 -2 V
28 -2 V
28 -1 V
28 -1 V
28 -1 V
29 0 V
28 0 V
28 1 V
29 1 V
28 1 V
28 1 V
27 2 V
28 3 V
27 2 V
27 3 V
27 4 V
27 4 V
26 4 V
26 4 V
26 5 V
25 5 V
25 6 V
24 6 V
25 7 V
24 6 V
23 8 V
23 7 V
23 8 V
23 9 V
22 9 V
22 9 V
22 10 V
22 10 V
21 11 V
22 11 V
21 12 V
20 12 V
21 13 V
21 13 V
20 14 V
21 14 V
20 15 V
20 15 V
21 16 V
20 17 V
20 17 V
20 18 V
21 18 V
20 20 V
21 19 V
20 21 V
21 21 V
21 22 V
20 23 V
21 23 V
22 25 V
21 25 V
21 26 V
22 27 V
22 28 V
22 29 V
23 29 V
22 31 V
23 32 V
23 33 V
24 34 V
23 35 V
24 36 V
25 38 V
24 38 V
25 40 V
25 42 V
26 42 V
26 44 V
26 46 V
27 47 V
27 48 V
27 51 V
28 51 V
28 54 V
1.000 UL
LT0
150 100 M
3300 0 V
1.000 UL
LT0
975 2060 M
975 100 L
1.000 UL
LTb
1.000 UL
LT2
150 100 M
33 0 V
34 0 V
33 0 V
33 0 V
34 0 V
33 0 V
33 0 V
34 0 V
33 0 V
33 0 V
34 0 V
33 0 V
33 0 V
34 0 V
33 0 V
33 0 V
34 0 V
33 0 V
33 0 V
34 0 V
33 0 V
33 0 V
34 0 V
33 0 V
33 0 V
34 2 V
33 4 V
33 7 V
34 9 V
33 12 V
33 14 V
34 17 V
33 19 V
33 21 V
34 24 V
33 27 V
33 29 V
34 31 V
33 34 V
33 36 V
34 39 V
33 41 V
33 44 V
34 46 V
33 49 V
33 51 V
34 53 V
33 56 V
33 59 V
34 61 V
33 63 V
33 66 V
34 68 V
33 71 V
33 73 V
34 76 V
33 78 V
33 81 V
34 83 V
33 85 V
33 88 V
34 91 V
33 93 V
33 95 V
22 64 V
1.000 UL
LT0
1483 100 M
0 286 V
34 39 V
33 41 V
33 44 V
34 46 V
33 49 V
33 51 V
34 53 V
33 56 V
33 59 V
34 61 V
33 63 V
33 66 V
34 68 V
33 71 V
33 73 V
34 76 V
33 78 V
33 81 V
34 83 V
33 85 V
33 88 V
34 91 V
33 93 V
33 95 V
22 64 V
1.000 UL
LTb
1.000 UL
LT3
333 644 M
24 15 V
25 15 V
24 16 V
26 18 V
26 18 V
26 19 V
26 20 V
27 21 V
27 21 V
28 21 V
28 21 V
29 21 V
29 21 V
29 19 V
30 19 V
30 17 V
30 15 V
30 14 V
31 11 V
31 9 V
31 6 V
31 3 V
31 1 V
31 -2 V
31 -5 V
31 -7 V
31 -10 V
31 -12 V
30 -15 V
30 -16 V
30 -18 V
30 -19 V
29 -20 V
28 -20 V
29 -21 V
28 -22 V
27 -21 V
27 -21 V
27 -20 V
27 -20 V
25 -19 V
26 -18 V
25 -16 V
25 -16 V
24 -15 V
25 -14 V
23 -12 V
24 -12 V
23 -10 V
24 -9 V
23 -7 V
22 -7 V
23 -6 V
22 -4 V
23 -3 V
22 -3 V
23 -1 V
22 -1 V
22 1 V
22 1 V
23 3 V
22 3 V
23 4 V
22 5 V
23 6 V
23 7 V
23 7 V
23 9 V
24 9 V
23 10 V
24 11 V
25 11 V
24 13 V
25 13 V
25 14 V
25 15 V
26 16 V
26 17 V
26 18 V
27 18 V
27 20 V
27 20 V
28 21 V
28 23 V
29 23 V
29 24 V
30 26 V
30 26 V
30 28 V
31 29 V
32 30 V
32 31 V
33 32 V
33 34 V
33 35 V
35 37 V
35 38 V
35 39 V
36 41 V
stroke
grestore
end
showpage
}}%
\put(2600,1876){\makebox(0,0)[l]{$e^{2{\Phi}(R)}B(R)\;\; (\kappa \ne0)$}}%
\put(2918,1420){\makebox(0,0)[l]{$B(R) \;\; (\kappa \ne 0) $}}%
\put(2000,2100){\makebox(0,0)[l]{$B(R) \;\; (\kappa = 0) $}}%
\put(546,632){\makebox(0,0)[l]{$\gamma=1$}}%
\put(1800,436){\makebox(0,0)[l]{$\gamma=-1$}}%
\put(1165,688){\makebox(0,0)[l]{$\gamma=0$}}%
\put(975,34){\makebox(0,0)[r]{$0$}}%
\put(1505,34){\makebox(0,0)[r]{$1$}}%
\put(3730,100){\makebox(0,0)[r]{$R$}}%
\end{picture}%
\endgroup
 